\documentclass{article}
\usepackage{amsmath,accents}
\usepackage{url}
\usepackage[colorlinks=cyan,hyperfootnotes=true,citecolor=cyan]{hyperref}
\usepackage{eufrak}
\usepackage{amsfonts}
\usepackage{amssymb}
\usepackage{amsthm}
\usepackage{color}

\usepackage{tikz}
\usepackage{tikzsymbols}
\usepackage[utf8]{inputenc} 
\usepackage{times}
\usepackage{mathtools}
\usepackage{footnote}
\makesavenoteenv{table}

\usetikzlibrary{arrows,shapes}
\usetikzlibrary{matrix}
\usetikzlibrary{shadows}
\usetikzlibrary{positioning}

\newcommand{\be}{\begin{equation}}
\newcommand{\ee}{\end{equation}}
\newcommand{\bea}{\begin{eqnarray}}
\newcommand{\eea}{\end{eqnarray}}

\begin{document}

\title{\bf Geometry and covariance of\\ symmetric teleparallel theories of gravity}

\author{Daniel Blixt${}^{1}$, Alexey Golovnev${}^{2}$, Maria-Jose Guzman${}^{3}$, Ramazan Maksyutov${}^{4,5}$\\
{\small ${}^{1}${\it Scuola Superiore Meridionale} }\\
{\small\it{Largo S. Marcellino 10, I-80138, Napoli, Italy} }\\
{\small d.blixt@ssmeridionale.it}\\
{\small ${}^{2}${\it Centre for Theoretical Physics, the British University in Egypt,}}\\ 
{\small {\it BUE 11837, El Sherouk City, Cairo Governorate, Egypt}}\\
{\small agolovnev@yandex.ru}\\
{\small ${}^{3}${\it Laboratory of Theoretical Physics, Institute of Physics, University of Tartu,}}\\ 
{\small {\it W. Ostwaldi 1, 50411 Tartu, Estonia}}\\
{\small mjguzman@ut.ee}\\
{\small ${}^{4}${\it Department of High Energy and Elementary Particle Physics, Saint Petersburg State University,}}\\ 
{\small {\it Ulianovskaya, 1, Stary Petergof,  Saint Petersburg, Russia}}\\
{\small ${}^{5}${\it Petersburg Nuclear Physics Institute of National Research Centre “Kurchatov Institute”,}}\\ 
{\small {\it Gatchina, 188300, Russia}}\\
{\small RamazanMaksyutov@yandex.ru}
}
\date{}

\maketitle

\begin{abstract}

We present the geometric foundations and derivations of equations of motion for symmetric teleparallel theories of gravity in the coincident gauge and covariant frameworks. We discuss the theoretical challenges introduced by the auxiliary fields responsible for the covariantisation procedure. We elucidate a tetradic structure interpretation behind this covariant formulation. Regarding the effect of covariantisation at the level of the equations of motion, we explicitly show that the only physical change, in case of setting an arbitrary energy-momentum tensor to the right hand side, resides in the requirement of the fulfillment of the covariant conservation laws. Also, we have explicitly introduced the fundamental covariantly-conserved teleparallel tetrad for the symmetric teleparallel frameworks. 

\end{abstract}

\section{Introduction}

Current theoretical research is actively engaged in exploring  modifications of general relativity (GR). While traditional modifications rely on curvature-based approaches like $f(\mathbb R)$ gravity and others, alternative geometric foundations have recently gained significant attention. In particular, there is intensive activity in gravitational theories based on the teleparallel framework, also denoted as metric teleparallel,  which describe gravity in terms of the torsion of a metric affine connection \cite{Aldrovandi:2013wha,Golovnev:2018red}. A less explored, but equally interesting option of modifying gravity is the symmetric teleparallel approach which instead makes use of a curvatureless and torsionless connection with nonvanishing non-metricity \cite{BeltranJimenez:2017tkd,BeltranJimenez:2018vdo,Gomes:2022vrc}.

In this paper, we will give a general introduction to and make some new research steps in the symmetric teleparallel framework. This approach encompasses the Symmetric Teleparallel Equivalent of General Relativity (STEGR), which is a formulation alternative to the Einstein-Hilbert action for gravity regarding the non-metricity scalar built up from the non-metricity tensor. The form of the action of STEGR in the so-called coincident gauge is identical to the usual expression quadratic in the connection, that is, the historical $\Gamma\Gamma$ action of Einstein. It lacks proper covariance under diffeomorphisms, unlike the elegant action found by Hilbert that includes the partial derivatives of the connection. Therefore, the simple generalisations of the STEGR action break diffeomorphisms, the same way as local Lorentz invariance gets broken in modified metric teleparallel models \cite{Golovnev:2020zpv}.

Given this break-down of diffeomorphisms, a natural question arises of how to formally restore such symmetry in modified symmetric teleparallel models, similar to what had been done in metric teleparallelism \cite{Golovnev:2017dox}. This can actually be implemented by simply promoting the coordinates which correspond to zero connection coefficients to a set of scalar fields on the manifold, and then allowing for otherwise arbitrary coordinates. Of course, it is nothing but a formal St{\" u}ckelberg procedure which does not remove the difference in the amount of symmetries between STEGR and modified models, the same as it was in the case of the teleparallel equivalent of general relativity (TEGR) and its modifications.

At this stage, we find a controversial issue concerning the covariantisation procedure, which regards the initial model as a gauge-fixed version of the covariantised one. It is well-known that gauge-fixing directly inside the action might give undesired outcomes.  However, in the metric teleparallel framework, at least if no matter directly couples to the spin connection, the covariantised action does give equations of motion that are fully equivalent to those coming from the non-covariant approach \cite{Golovnev:2017dox, Golovnev:2023yla}, and consequently the same number of propagating degrees of freedom (dof) \cite{Golovnev:2021omn}. The trick is to rewrite the physical tetrad in terms of an arbitrary one, which is endowed with the same physical content \cite{Blixt:2022rpl}. Nonetheless, an important point  is that the gauge fixing is of a purely algebraic nature in the metric teleparallel.

In symmetric teleparallel, it is crucial to realise that the broken symmetry extends beyond a purely algebraic nature.  Moreover, covariant rewriting of modified symmetric teleparallel theories brings second-order derivatives into the action, in a non-trivial way. Fixing such gauges right inside the action is not always an innocent deed, with many worrisome examples coming from the diffeomorphisms themselves. As an example, let us imagine fixing the lapse and shift, $N=1$ and $N_i=0$, directly inside the ADM action, a choice that would be consistent with the primary constraints $\pi_{0\mu}$, while then missing the presence of the Hamiltonian and momenta constraints (which are secondary constraints). In other words, gauge fixing too early does not allow diffeomorphism symmetries to ``hit twice'' \cite{Golovnev:2022rui} and the counting of dof would end up with 6 instead of 2 dof. 

On the other hand, the process of covariantisation can be understood as a St{\" u}ckelberg procedure which in principle should not change the physical content of the model. In numerous papers the covariant symmetric teleparallel theories are taken as just a fully equivalent rewriting of the coincident-gauge ones, without rigorously proving this statement. At the same time, some other works do treat modified covariant STEGR and modified $\Gamma\Gamma$ gravity as different models \cite{BoJe, Boehmer:2023fyl}. All in all, the matter at hand seems rather non-trivial, with no consensus in the literature. Consequently, this point urgently needs clarification.

Our aim in this paper is to exhibit the structure of symmetric teleparallel theories, to explain the geometry behind, and to show how the proposed covariantisation \cite{BeltranJimenez:2022azb} actually works. One important message we have to convey is that the covariantisation fields $\xi^{\mu}(x)$ cannot be taken as components of a vector. To the contrary, they must be treated as a set of scalars, for they represent a possible choice of coordinates. Based on that, we also expose the fundamental (non-orthonormal) tetrads, i.e. the sets of covariantly constant vectors for symmetric teleparallel frameworks. This is an important result which provides the symmetric teleparallel models with a description similar to the pure-tetrad approach to metric teleparallel ones.

As to the effects of covariantisation, we explicitly show, both in general and in simple examples, that the equations of motion in vacuum are not modified, while in case of coupling to matter in terms of an arbitrary energy-momentum tensor, the covariantised theory only requires it to be covariantly conserved. This raises a challenging  question of whether the covariantisation is worth pursuing. On one hand, for covariantly conserved types of matter, it finally brings no change. Therefore, it only prohibits using non-conserved matter content which is possible in non-covariant modified symmetric teleparallel models. 

At the same time, one more price to pay for the covariant approach is an action principle with second derivatives of the additional fields $\xi^{\mu}$, and therefore fourth order equations of motion. A potential Hamiltonian description of the modified covariantised theory requires then to utilise the Ostrogradski procedure, or alternatively to introduce Lagrange multipliers. In comparison with modified metric teleparallel gravity \cite{Golovnev:2021omn}, it is expected to be significantly harder to fully analyse even the set of primary constraints. An important feature of the fundamental tetrads we propose is that they allow one to avoid the higher derivatives by explicitly constraining torsion to be zero.

We structure this manuscript as follows. We prepare our conventions and the mathematical background for the symmetric teleparallel geometry in Section \ref{sec:symmtel}. We present the derivation of the equations of motion for a general model in Section \ref{sec:EoM}. The covariantisation procedure is introduced in Section \ref{sec:cov}, and then we present examples of its usage in a toy model in Section \ref{sec:toy}. We present a similar analysis on the well-known $f(\mathbb Q)$ model in Section \ref{sec:fq}. Some discussion concerning the implications of the covariant procedure is presented in Section \ref{sec:dis}. We end up with the conclusions in Section \ref{sec:ccl}.

\section{Symmetric teleparallel framework}
\label{sec:symmtel}

In order to work in a symmetric teleparallel geometry, we will consider a manifold endowed with a metric structure given by the metric tensor $g_{\mu\nu}$ and an affine connection $\Gamma^{\alpha}{}_{\mu\nu}$ defining the notion of parallel transport along the manifold. Then the underlying geometry will be determined by the properties of the connection, which we will consider to have nontrivial non-metricity only. This means that, in particular, the torsion tensor is zero, i.e. the connection is symmetric, in addition to the requirement that the Riemann tensor is zero which defines the notion of teleparallel structure. In this way, the connection describing the symmetric teleparallel geometry takes the form 
\begin{equation}
\Gamma^{\alpha}{}_{\mu\nu}=\mathring{\Gamma}^{\alpha}{}_{\mu\nu}+L^{\alpha}_{\hphantom{\alpha}\mu\nu}.
\label{eq:symconnect}
\end{equation}
In this equation the usual GR connection (Levi-Civita) is $\mathring{\Gamma}^{\alpha}{}_{\mu\nu}$, and all the other quantities related to it  will also be denoted by a ring on top of them. The disformation tensor $L^{\alpha}_{\hphantom{\alpha}\mu\nu}$ in Eq. \eqref{eq:symconnect} is given by 
\be
L_{\alpha\mu\nu}=\frac12 \left(Q_{\alpha\mu\nu}-Q_{\mu\alpha\nu}-Q_{\nu\alpha\mu}\right)
\ee
in terms of the non-metricity tensor 
$$
Q_{\alpha\mu\nu}\equiv \nabla_{\alpha} g_{\mu\nu}.
$$

Using the formula \eqref{eq:symconnect} relating the two connections, it is easy to derive the relationship among their curvature tensors. In particular, if the connection is not only symmetric but also flat, denoting the Ricci scalars of $\Gamma$ and $\mathring{\Gamma}$ by ${\mathbb R}$ and $\mathring{\mathbb R}$, respectively, we find that 
\be
0={\mathbb R}=\mathring{\mathbb R}+\mathbb Q + \mathbb B
\label{telecondition}
\ee 
where the following definitions have been made
\be
{\mathbb Q}=\frac14 Q_{\alpha\mu\nu}Q^{\alpha\mu\nu}-\frac12 Q_{\alpha\mu\nu}Q^{\mu\alpha\nu}-\frac14 Q_{\mu}Q^{\mu}+\frac12 Q_{\mu}\tilde Q^{\mu},
\ee
\be
{\mathbb B}=g^{\mu\nu}\mathring{\nabla}_{\alpha}L^{\alpha}_{\hphantom{\alpha}\mu\nu}-\mathring{\nabla}^{\beta}L^{\alpha}_{\hphantom{\alpha}\alpha\beta}=\mathring{\nabla}_{\alpha}\left(Q^{\alpha}-\tilde Q^{\alpha}\right),
\ee
with the non-metricity traces defined as
$$Q_{\alpha}\equiv Q^{\hphantom{\alpha}\mu}_{\alpha\hphantom{\mu}\mu}\qquad \mathrm{and} \qquad\tilde Q_{\alpha}\equiv Q^{\mu}_{\hphantom{\mu}\mu\alpha}.$$
The so-called non-metricity scalar $\mathbb Q$ can also be written as
$${\mathbb Q}=\frac12 P^{\alpha\mu\nu}Q_{\alpha\mu\nu},$$
with a superpotential defined as
\be
P^{\alpha\mu\nu}=L^{\alpha\mu\nu}+\frac12 g^{\mu\nu}(\tilde Q^{\alpha}-Q^{\alpha})+\frac14 (g^{\alpha\mu}Q^{\nu}+g^{\alpha\nu}Q^{\mu}).
\label{SPot}
\ee
This superpotential, or the non-metricity conjugate, is the derivative of the nonmetricity scalar with respect to the nonmetricity tensor, $P^{\alpha\mu\nu}=\frac{\delta \mathbb Q}{\delta Q_{\alpha\mu\nu}}$.

Therefore, from the Eq.  \eqref{telecondition} we see that if the symmetric connection is flat, then the scalar $\mathbb Q$ is different from $-\mathring{\mathbb R}$ only by a boundary term, thus producing a model equivalent to GR at the level of the action. As a matter of fact, the simplest flat symmetric connection is given by $\Gamma^{\alpha}{}_{\mu\nu}=0$ in which case $Q_{\alpha\mu\nu}\equiv \partial_{\alpha} g_{\mu\nu}$, and the action
\be 
S=\int d^4 x \sqrt{-g}\cdot \mathbb Q
\ee 
is nothing but Einstein's $\Gamma\Gamma$ action, that is the part of the Ricci scalar quadratic in the Levi-civita connection. This is the action for STEGR in the so-called coincident gauge. Any different choice of this type of connection can, at least locally, be transformed to the coincident gauge, that is zero, by a coordinate change. 

The STEGR action can be used as the starting point for building alternative gravity theories, which we will consider in the following, in the coincident gauge. However, it is also possible to build such models in the symmetric teleparallel framework for a general symmetric teleparallel connection, which will be presented in Sec. \ref{sec:cov}.

\subsection{Generalisations of STEGR}

The breaking of diffeomorphism symmetry in STEGR is harmless enough since it was restricted to the boundary term $\mathbb B$, therefore not influencing the equations of motion. The same way as it happens in TEGR,  we can talk about STEGR as a diffeomorphism pseudo-invariant theory. However, the story is different when considering nonlinear modified models based on the STEGR Lagrangian. One of the most important such models is $f(\mathbb Q)$ gravity given by the action
$$S=\int d^4 x \sqrt{-g}\cdot f(\mathbb Q).$$
The would-be boundary term is no longer just a boundary term there, unless the function $f$ is linear. We will discuss in the next sections the consequences of this peculiar property.

A less trivial modification, called Newer GR, takes an alternative quadratic form of the non-metricity tensor:
\be
\label{EEE}
{\mathfrak Q}\equiv \frac12 {\mathcal E}^{ \alpha_1\mu_1\nu_1\alpha_2\mu_2\nu_2}_{ \rho_1\sigma_1\rho_2\sigma_2\rho_3\sigma_3} g^{\rho_1\sigma_1} g^{\rho_2\sigma_2} g^{\rho_3\sigma_3} Q_{\alpha_1\mu_1\nu_1} Q_{\alpha_2\mu_2\nu_2},
\ee
with an arbitrary tensor ${\mathcal E}^{ \alpha_1\mu_1\nu_1\alpha_2\mu_2\nu_2}_{ \rho_1\sigma_1\rho_2\sigma_2\rho_3\sigma_3}$ constructed purely from Kronecker  $\delta^{\mu}_{\nu}$-s. Note that, in this way, we do not consider parity-odd terms. This form has some resemblance with the premetric program for TEGR, see for instance Refs. \cite{Ferraro:2016wht,Itin:2016nxk,Guzman:2020kgh}. We present the most general non-metricity scalar here in this uncommon form for giving a general gist of the model, but also for exhibiting in a compact way the derivation of the equations of motion in the beginning of the next Section. For the convenience of that, we assume that all the natural symmetries of this tensor are satisfied. Namely, as is natural from the formula \eqref{EEE}, we demand that the following permutations leave the $\mathcal E$ tensor unchanged: $\mu_i\longleftrightarrow\nu_i$, $\rho_i\longleftrightarrow\sigma_i$, $(\rho\sigma)_i\longleftrightarrow(\rho\sigma)_j$, and $(\alpha\mu\nu)_1\longleftrightarrow(\alpha\mu\nu)_2$. In particular, modulo different factors of $\frac12$ in the definitions, the usual non-metricity conjugate \cite{BeltranJimenez:2019tme} is given by
\be
{\mathfrak P}^{\alpha\mu\nu}= {\mathcal E}^{ \alpha\mu\nu\alpha_2\mu_2\nu_2}_{ \rho_1\sigma_1\rho_2\sigma_2\rho_3\sigma_3} g^{\rho_1\sigma_1} g^{\rho_2\sigma_2} g^{\rho_3\sigma_3} Q_{\alpha_2\mu_2\nu_2}.
\ee
Unfortunately, in any concrete example, it would be quite cumbersome to explicitly symmetrise the $\mathcal E$ tensor.

Despite the mentioned difficulties of the total symmetrisation, it still makes sense to explicitly keep the symmetry among the two non-metricity tensors while treating the three inverse metrics individually, one by one. To this end, we take the arbitrary scalar
\begin{equation}
\label{arbqscal}
2{\mathfrak Q}=c_1 Q_{\alpha\mu\nu}Q^{\alpha\mu\nu}+c_2 Q_{\alpha\mu\nu}Q^{\mu\alpha\nu} +c_3 Q_{\mu}Q^{\mu} + c_4 {\tilde Q}_{\mu}\tilde Q^{\mu}+c_5 Q_{\mu}\tilde Q^{\mu}
\end{equation}
which defines the so-called Newer GR. Then we rewrite it as 
$${\mathfrak Q}=\frac12 {\mathfrak P}^{\alpha\mu\nu} Q_{\alpha\mu\nu},$$
with the non-metricity conjugate given then by
\bea 
\label{arbqconj}
{\mathfrak P}^{\alpha\mu\nu} & = & c_1 Q^{\alpha\mu\nu} + \frac{c_2}{2} \left(Q^{\mu\nu\alpha} + Q^{\nu\mu\alpha}\right) +c_3 Q^{\alpha} g^{\mu\nu} +\frac{c_4}{2}\left(g^{\alpha\mu}{\tilde Q}^{\nu} + g^{\alpha\nu}{\tilde Q}^{\mu}\right) \nonumber \\
& &  +\frac{c_5}{4}\left(2{\tilde Q}^{\alpha} g^{\mu\nu} + g^{\alpha\mu}Q^{\nu} + g^{\alpha\nu}Q^{\mu}\right).
\eea
Newer GR models have been introduced in Ref. \cite{BeltranJimenez:2017tkd}, and although they propagate more dof than GR, they have the limit of STEGR and are a starting point to explore modified gravity based on the symmetric teleparallel framework. 

\section{Equations of motion}
\label{sec:EoM}

In this Section we will present the equations of motion for the modified symmetric teleparallel model given in terms of an arbitrary non-metricity scalar (\ref{arbqscal}), or equivalently a non-metricity conjugate (\ref{arbqconj}).  We are still considering the coincident gauge, therefore $Q_{\alpha\mu\nu}\equiv \partial_{\alpha} g_{\mu\nu}$.

As a warm-up, note that, for a general non-linear theory
\begin{equation}
\label{genact}
S=\int d^4 x \sqrt{-g}\cdot f(\mathfrak Q),
\end{equation}
 taken in the form of \eqref{EEE}, the variation of the tensor $\delta \mathcal E =0$ vanishes because it has purely constant components. Consequently, the equations of motion, from the $\delta g_{\mu\nu}$ variation of the action \eqref{genact} multiplied by $-2$, look very simple:
\begin{multline*}
2\partial_{\alpha} \left(\sqrt{-g}f^{\prime}\cdot {\mathcal E}^{ \alpha\mu\nu\alpha_2\mu_2\nu_2}_{ \rho_1\sigma_1\rho_2\sigma_2\rho_3\sigma_3} g^{\rho_1\sigma_1} g^{\rho_2\sigma_2} g^{\rho_3\sigma_3} Q_{\alpha_2\mu_2\nu_2} \right)\\ +\sqrt{-g} \left(3f^{\prime} \cdot {\mathcal E}^{ \alpha_1\mu_1\nu_1\alpha_2\mu_2\nu_2}_{ \rho_1\sigma_1\rho_2\sigma_2\rho_3\sigma_3} g^{\rho_1\mu} g^{\sigma_1\nu} g^{\rho_2\sigma_2} g^{\rho_3\sigma_3} Q_{\alpha_1\mu_1\nu_1} Q_{\alpha_2\mu_2\nu_2} - f g^{\mu\nu}\right)=0,
\end{multline*}
with the trouble mentioned above: it would be quite hard to explicitly have the tensor $\mathcal E$ fully symmetrised.

Actually, working in the standard representation of the formulae (\ref{arbqscal}) or (\ref{arbqconj}), the variations can also be done quite nicely. For the sake of simplicity, let us start from the Newer GR, i.e. a unity function $f$ in the action:
$$S=\int d^4 x \sqrt{-g}\cdot\mathfrak Q.$$
By a direct inspection, term by term, one can easily see that
\be
Q^{\alpha\mu\nu}\cdot \delta {\mathfrak P}_{\alpha\mu\nu}= {\mathfrak P}^{\alpha\mu\nu}\cdot \delta Q_{\alpha\mu\nu}
\ee
from what it immediately follows that
\be
\delta ( {\mathfrak P}^{\alpha\mu\nu} Q_{\alpha\mu\nu}) = 2 {\mathfrak P}^{\alpha\mu\nu} \cdot \partial_{\alpha} \delta g_{\mu\nu} - \left({\mathfrak P}^{\mu\alpha\beta} {Q^{\nu}}_{\alpha\beta} +2{\mathfrak P}^{\alpha\beta\mu}{Q_{\alpha\beta}}^{\nu}  \right)\cdot \delta g_{\mu\nu}.
\ee
One can also note, by a direct calculation, that ${\mathfrak P}^{\mu\alpha\beta} {Q^{\nu}}_{\alpha\beta} +2{\mathfrak P}^{\alpha\beta\mu}{Q_{\alpha\beta}}^{\nu} $ is automatically $\mu\leftrightarrow\nu$ symmetric, so that no more symmetrisation is required, and the equation obtained by varying the action can be written as
\begin{equation}
\label{NGRindup}
\frac{2}{\sqrt{-g}}\partial_{\alpha}\left(\sqrt{-g} {\mathfrak P}^{\alpha\mu\nu}\right) + {\mathfrak P}^{\mu\alpha\beta} {Q^{\nu}}_{\alpha\beta} +2{\mathfrak P}^{\alpha\beta\mu}{Q_{\alpha\beta}}^{\nu}  - {\mathfrak Q} g^{\mu\nu}=0,
\end{equation}
upon having multiplied it by $-2$.

By lowering the indices under the derivative sign, we can rewrite the equation with upper indices (\ref{NGRindup}) as
\begin{equation}
\label{NGRindmix}
\frac{2}{\sqrt{-g}}\partial_{\alpha}\left(\sqrt{-g} {{\mathfrak P}^{\alpha\mu}}_{\nu}\right) + {\mathfrak P}^{\mu\alpha\beta} Q_{\nu\alpha\beta} - {\mathfrak Q} \delta^{\mu}_{\nu}=0
\end{equation}
with the mixed position of indices, or
\begin{equation}
\label{NGRinddown}
\frac{2}{\sqrt{-g}}\partial_{\alpha}\left(\sqrt{-g} {{\mathfrak P}^{\alpha}}_{\mu\nu}\right) +{{\mathfrak P}_{\mu}}^{\alpha\beta} Q_{\nu\alpha\beta}- 2{\mathfrak P}_{\alpha\beta\nu}{Q^{\alpha\beta}}_{\mu} - {\mathfrak Q} g_{\mu\nu}=0
\end{equation}
with all the indices down. It is worth noticing that  the expression of $ {{\mathfrak P}_{\mu}}^{\alpha\beta} Q_{\nu\alpha\beta}- 2{\mathfrak P}_{\alpha\beta\nu}{Q^{\alpha\beta}}_{\mu}$ in the Eq. (\ref{NGRinddown})  corresponds to the $q_{\mu\nu}$ tensor of the previous works \cite{BeltranJimenez:2019tme}. At the same time, the tensor with the mixed position of indices (\ref{NGRindmix}), which can also be found there \cite{BeltranJimenez:2019tme}, though obviously not symmetric in itself, takes the simple form of
\be 
{\mathfrak P}^{\mu\alpha\beta} Q_{\nu\alpha\beta}= c_1  Q^{\mu\alpha\beta}Q_{\nu\alpha\beta} +c_2 Q^{\alpha\beta\mu} Q_{\nu\alpha\beta} + c_3 Q^{\mu}Q_{\nu} + c_4 {Q_{\nu}}^{\mu\alpha} {\tilde Q}_{\alpha} + \frac{c_5}{2} \left({\tilde Q}^{\mu}Q_{\nu} + {Q_{\nu}}^{\mu\alpha} Q_{\alpha}\right).
\ee 

The generalisation of the equations of motion for the general $f(\mathfrak Q)$ theory with the action given by \eqref{genact} is then straightforward:
\begin{equation}
\label{fqcoin}
\frac{2}{\sqrt{-g}}\partial_{\alpha}\left(\sqrt{-g} f^{\prime}({\mathfrak Q}) {{\mathfrak P}^{\alpha\mu}}_{\nu}\right) +  f^{\prime}({\mathfrak Q}) {\mathfrak P}^{\mu\alpha\beta} Q_{\nu\alpha\beta} -  f({\mathfrak Q}) \delta^{\mu}_{\nu}=0,
\end{equation}
or equivalently, and in purely tensorial quantities when substituting $\partial_{\alpha}$ by $\nabla_{\alpha}$,
\begin{equation}
\label{fqcoincov}
2\partial_{\alpha}\left( f^{\prime}({\mathfrak Q}) {{\mathfrak P}^{\alpha\mu}}_{\nu}\right) +  f^{\prime}({\mathfrak Q}) \left(Q_{\alpha}{{\mathfrak P}^{\alpha\mu}}_{\nu}+{\mathfrak P}^{\mu\alpha\beta} Q_{\nu\alpha\beta}\right) -  f({\mathfrak Q}) \delta^{\mu}_{\nu}=0,
\end{equation}
while the matter can be put as $-2T^{\mu}_{\nu}$ to the right hand side, assuming the usual fundamental constants combination of $\frac{8\pi G}{c^4}=1$.

\subsection{An independent check}

As an independent check of the scheme developed above, we can do all the variations explicitly term by term:
\begin{eqnarray}
\delta(Q_{\alpha\mu\nu}Q^{\alpha\mu\nu}) & = & 2Q^{\alpha\mu\nu}\cdot \partial_{\alpha}\delta g_{\mu\nu} - (Q^{\mu\alpha\beta}{Q^{\nu}}_{\alpha\beta}+ 2Q^{\alpha\beta\mu}{Q_{\alpha\beta}}^{\nu})\cdot \delta g_{\mu\nu},\\
\delta(Q_{\alpha\mu\nu}Q^{\mu\alpha\nu}) & = & (Q^{\mu\alpha\nu}+Q^{\nu\alpha\mu})\cdot \partial_{\alpha}\delta g_{\mu\nu} - (Q^{\mu\alpha\beta}{Q_{\alpha\beta}}^{\nu}+ Q^{\nu\alpha\beta}{Q_{\alpha\beta}}^{\mu} + Q^{\alpha\beta\mu}{Q_{\beta\alpha}}^{\nu})\cdot \delta g_{\mu\nu},\\
\delta(Q_{\alpha}Q^{\alpha}) & = & 2Q^{\alpha}g^{\mu\nu}\cdot \partial_{\alpha}\delta g_{\mu\nu} - (Q^{\mu}Q^{\nu}+ 2Q^{\alpha\mu\nu}Q_{\alpha})\cdot \delta g_{\mu\nu},\\
\delta({\tilde Q}_{\alpha}{\tilde Q}^{\alpha}) & = & ({\tilde Q}^{\mu}g^{\alpha\nu} + {\tilde Q}^{\nu}g^{\alpha\mu})\cdot \partial_{\alpha}\delta g_{\mu\nu} - ({\tilde Q}^{\mu}{\tilde Q}^{\nu}+ (Q^{\mu\nu\alpha}+Q^{\nu\mu\alpha}){\tilde Q}_{\alpha})\cdot \delta g_{\mu\nu}, \\
\delta({\tilde Q}_{\alpha} Q^{\alpha}) & = & \left({\tilde Q}^{\alpha}g^{\mu\nu} + \frac12 (Q^{\mu}g^{\alpha\nu}+ Q^{\nu}g^{\alpha \mu})\right)\cdot \partial_{\alpha}\delta g_{\mu\nu} \nonumber \\
& & - \left(Q^{\alpha\mu\nu}{\tilde Q}_{\alpha}+ \frac12 (Q^{\mu\nu\alpha}+Q^{\nu\mu\alpha})Q_{\alpha}+\frac12 ({\tilde Q}^{\mu}Q^{\nu}+{\tilde Q}^{\nu} Q^{\mu})\right)\cdot \delta g_{\mu\nu},
\end{eqnarray}
with all the necessary symmetrisations explicitly imposed. One can easily see that the brackets in front of $\delta g_{\mu\nu}$, with the proper $c_i$ coefficients, do sum up precisely to ${\mathfrak P}^{\mu\alpha\beta} {Q^{\nu}}_{\alpha\beta} +2{\mathfrak P}^{\alpha\beta\mu}{Q_{\alpha\beta}}^{\nu} $ in the equation (\ref{NGRindup}).

Another way to look at this result is from the representation with the $\mathcal E$-tensor  
\eqref{EEE}. Basically, the three terms, ${\mathfrak P}^{\mu\alpha\beta} {Q^{\nu}}_{\alpha\beta}$, ${\mathfrak P}^{\alpha\beta\mu}{Q_{\alpha\beta}}^{\nu} $, and ${\mathfrak P}^{\alpha\mu\beta}{{Q_{\alpha}}^{\nu}}_{\beta} $, come from the variations of the three inverse metrics. We sum over these three terms, which would be taken care of just by a factor of $3$ in case of an explicitly symmetrised $\mathcal E$-tensor. The non-trivial aspect is that, in this shape, it is automatically symmetric under the exchange $\mu\longleftrightarrow\nu$, for what the precise shape of the non-metricity conjugate (\ref{arbqconj}) is very important.

\subsection{The case of STEGR and $f(\mathbb Q)$}

It is also interesting to check that in the case of STEGR, $c_1=\frac12$, $c_2=-1$, $c_3=-\frac12$, $c_4=0$, $c_5=1$, the equations of motion reduce to the usual Einstein ones. For that, one can take
$$L^{\alpha}_{\hphantom{\alpha}\mu\nu}= \frac12 \left(Q^{\alpha}_{\hphantom{\alpha}\mu\nu} - Q_{\mu\nu}{}^{\alpha} -  Q_{\nu\mu}{}^{\alpha}\right)  $$
and calculate, from $\Gamma= \mathring{\Gamma} + L $ and $R(\Gamma)=0$,
$$\mathring{R}^{\mu}_{\nu}=-\mathring{\nabla}_{\alpha} L^{\alpha\mu}{}_{\nu}+\mathring{\nabla}_{\nu} L_{\alpha}{}^{\alpha\mu} - L^{\alpha}{}_{\alpha\rho} L^{\rho\mu}{}_{\nu} + L^{\alpha\beta\mu} L_{\beta\alpha\nu}.$$
In passing, note that using $L^{\alpha}{}_{\alpha\mu}=-\frac12 Q_{\mu}$ an important symmetry property follows: 
\begin{equation}
\label{symvec}
\mathring{\nabla}_{\nu} Q_{\mu} = \mathring{\nabla}_{\mu} Q_{\nu}
\end{equation}
which can also be checked directly and is also valid without taking the coincident gauge.

If we take the trace of $\mathring{R}^{\mu}_{\nu}$, we can easily find that actually $\mathring{R}=-{\mathbb Q} - {\mathbb B}$, with the quantities defined in the previous Section. A little bit more cumbersome a calculation, taking into account the symmetry property (\ref{symvec}) and an obvious relation
$$\frac{1}{\sqrt{-g}}\partial_{\alpha}\left(\sqrt{-g} {P^{\alpha\mu}}_{\nu}\right)= \mathring{\nabla}_{\alpha} {P^{\alpha\mu}}_{\nu} + \mathring{\Gamma}^{\rho}_{\alpha\nu} P^{\alpha\mu}{}_{\rho} - \mathring{\Gamma}^{\mu}_{\alpha\rho} P^{\alpha\rho}{}_{\nu}, $$
shows also that
\begin{equation}
\label{Ricciten}
\frac{1}{\sqrt{-g}}\partial_{\alpha}\left(\sqrt{-g} {P^{\alpha\mu}}_{\nu}\right) + \frac12 P^{\mu\alpha\beta}Q_{\nu\alpha\beta} = -\mathring{R}^{\mu}_{\nu} - \frac12 {\mathbb B} \delta^{\mu}_{\nu}.
\end{equation}
Therefore, the equation of motion (\ref{NGRindmix}) we have derived above in fact reduces to $-2\mathring{G}^{\mu}_{\nu}=0$ in the case of STEGR.

It allows to rewrite the equations (\ref{fqcoin}) for $f(\mathbb Q)$ in the following form (having divided those by $-2$):
\begin{equation}
\label{eqfq}
f^{\prime} \mathring{G}^{\mu}_{\nu} +\frac12 \left(f - f^{\prime} {\mathbb Q}\right) \delta^{\mu}_{\nu} - f^{\prime\prime} P^{\alpha\mu}{}_{\nu} \partial_{\alpha} {\mathbb Q} =0.
\end{equation}
In particular, we see that, analogously to the $f(\mathbb T)$ gravity, the constant $\mathbb Q$ solutions are just solutions of GR, with a cosmological constant, and renormalised gravitational constant.

It is evident that Eq. \eqref{eqfq} can also be derived directly from the $f(\mathbb Q)$ action by using the simple fact that ${\mathbb Q}= - \mathring{R} - \mathbb B$. For doing so, we would need the well-known variation
$$\delta \mathring{R} = - \mathring{R}^{\mu\nu} \delta g_{\mu\nu} + \left(\mathring{\nabla}^{\mu} \mathring{\nabla}^{\nu} - g^{\mu\nu} \mathring{\square} \right) \delta g_{\mu\nu}$$
as well as another formula
$$\delta {\mathbb B}= \left( g^{\mu\nu} \mathring{\square} - \mathring{\nabla}^{\mu} \mathring{\nabla}^{\nu} \right) \delta g_{\mu\nu} - \mathring{\nabla}_{\alpha} \left(P^{\alpha\mu\nu} \delta g_{\mu\nu} \right) - \frac12 {\mathbb B} g^{\mu\nu} \delta g_{\mu\nu} $$
which can be derived (in the coincident gauge) from implications of the relation $\mathring{\Gamma}^{\alpha}_{\mu\nu}= - L^{\alpha}{}_{\mu\nu}$ together with the elementary variational identity $\delta \Gamma^{\alpha}_{\mu\nu}=\frac12 g^{\alpha\beta} \left(\mathring{\nabla}_{\mu}\delta g_{\beta\nu} + \mathring{\nabla}_{\nu}\delta g_{\beta\mu} - \mathring{\nabla}_{\beta}\delta g_{\mu\nu}\right)$. It is worth acknowledging that pursuing this approach does not yield significant simplification compared to the aforementioned brute-force method. Nevertheless, this variation holds potential importance in the realm of more generalised symmetric teleparallel models of the $f(\mathbb{Q},\mathbb{B})$ type, analogous to $f(\mathbb T, \mathbb B)$ in metric teleparallelism.

\section{Covariantisation}
\label{sec:cov}

In this section we will introduce the geometric properties of the covariant approach in a symmetric teleparallel framework, the consistent implementation at the level of the equations of motion, and prepare for the following sections with examples of cosmology.

Naturally, in the case of coincident gauge, the symmetry under diffeomorphisms is not preserved. Specifically, when considering a symmetric teleparallel connection that objectively exists on the spacetime manifold, we can select coordinates in which it vanishes entirely. Nonetheless, the covariant approach defines the connection equal to
\begin{equation}
\label{theconn}
\Gamma^{\rho}{}_{\mu\nu}=  \frac{\partial^2 \xi^{\alpha} }{\partial x^{\mu} \partial x^{\nu} }\cdot \frac{\partial x^{\rho} }{\partial \xi^{\alpha} }.
\end{equation}
These are the connection coefficients that we would get by an arbitrary coordinate transformation from a frame with zero connection. Then $g_{\mu\nu}$ and $\xi^{\alpha}$ are taken as independent variables in the variations.

\subsection{The geometric features}

The connection given by Eq. \eqref{theconn} is evidently symmetric in the lower indices, and it can also be readily verified that it is flat. Furthermore, it is possible to represent every connection of this sort in this manner, at least locally. However, it is important to be accurate. It might seem enticing to transform $\xi^{\alpha}$ as a vector, but it is easy to see that in doing so the Eq. \eqref{theconn} would look far from being covariant, as an expression for a connection coefficient.

The appropriate transformation for the quantities $\xi^{\alpha}$ corresponds to treating them as a set of scalars. This is motivated by the following rationale: $\xi^{\alpha}$ represents the coordinates in which $\Gamma^{\rho}{}_{\mu\nu}=0$. In other words, the connection \eqref{theconn} in the coordinates $x$ is such as it should be if it was zero in coordinates $\xi(x)$. Consequently, we must allow ourselves the freedom of choosing our coordinates $x$, but it would be reasonable to just fix the coordinates $\xi$ in which $\Gamma=0$. Once this fixing has been established, the $\xi$ coordinates manifest as a collection of real-valued functions, specifically scalars, defined on the manifold.

Subsequently, it becomes evident that we have obtained a set of 1-forms that form a basis
\begin{equation}
\label{basicforms}
e^n_{\mu}\equiv\frac{\partial\xi^n}{\partial x^{\mu}},
\end{equation}
or a (co-)tetrad with zero spin-connection, and the spacetime connection coefficients (\ref{theconn}) are 
\be
\label{ourconn}
\Gamma^{\alpha}{}_{\mu\nu}=e^{\alpha}_n \partial_{\mu} e^n_{\nu}
\ee
with $e^{\alpha}_{n}$ being the matrix inverse of $e^n_{\alpha}$. The index of $\xi$ was changed to a Latin letter in order to stress that it is not a component of a vector, and it simply numbers the coordinates, or 1-forms of the tetrad. 

Note that here we use a more general notion of a tetrad than what is usually employed in GR. Namely, our tetrads are not assumed to be orthonormal. In the standard gravity, orthonormality is a natural and convenient requirement, also needed for coupling fermionic fields as a way of introducing the Lorentz group into the game. However, in general teleparallel frameworks, and as a description of tangent space bases, it is more natural to have arbitrary tetrads. Moreover, a coordinate basis as a tetrad (\ref{basicforms}) of the connection (\ref{ourconn}) would be orthonormal in the simple (pseudo)Euclidean spaces only.

Similar to the scenario in metric teleparallelism, we obtain that the set of 1-forms \eqref{basicforms} is covariantly conserved:  $$\nabla_{\alpha} e^n_{\beta}=\partial_{\alpha} e^n_{\beta} - \Gamma^{\gamma}{}_{\alpha\beta} e^n_{\gamma}=0,$$ 
and moreover, the transformation behaviour of the connection coefficients aligns genuinely with that of a connection. Unlike in the metric case, there is no anholonomy, that is $\partial_{\mu} e^n_{\nu}=\partial_{\nu} e^n_{\mu}$, and therefore no torsion. On the other hand, we can find again (compare with the Ref. \cite{LorGov}) that
\begin{equation}
\label{convar}
\delta\Gamma^{\alpha}{}_{\mu\nu} =e^{\alpha}_n \nabla_{\mu} \delta e^n_{\nu}=\nabla_{\mu} (e^{\alpha}_n  \delta e^n_{\nu}).
\end{equation}
This is how a variation of $\xi^{\alpha}(x)$ works in any symmetric teleparallel model.

Let us summarise the main geometric feature of every teleparallel gravity. A teleparallel geometry, i.e. zero curvature, means that there exists a basis of covariantly conserved vectors. It means $\nabla_{\mu}e^a_{\nu}=0$ in the sense of four independent 1-forms, or equivalently a soldering form which corresponds to zero spin connection. In metric teleparallel, the usual approach is that this tetrad as a dynamical variable is absolutely free (for sure, apart from non-degeneracy), while the metric is defined as $g_{\mu\nu}=\eta_{ab} e^a_{\mu} e^b_{\nu}$, so that an arbitrary tetrad is orthonormal by definition. In the symmetric teleparallel, the tetrad is holonomic, i.e. it is a basis of coordinate vectors $e^a_{\mu}=\frac{\partial\xi^a}{\partial x^{\mu}}$, while the metric is an independent variable.

These torsion-free covariantly constant tetrads (\ref{basicforms}) can also be viewed as an additional geometric structure allowing to rewrite a non-covariant notion in a covariant form, compare with Ref. \cite{unieq}. The situation of unimodular gravity discussed in that Reference is similar to modified symmetric teleparallel models. When we demand that $\sqrt{-g}=1$, it leaves only the traceless part of Einstein equations. However, as long as the matter is covariantly conserved by itself, it is fully equivalent to GR. There were some claims \cite{unipert} however that the cosmological perturbations are different from that of GR because the Newtonian gauge is not allowed. It is evident that such statements are not reasonable, for otherwise the physical GR predictions would depend on the choice of gauge. The paper \cite{unieq} elucidates the situation in yet another way by introducing a fixed fiducial measure which then allows to work in any gauge. Analogously, we also break the diffeomorphisms down, but it can be equivalently described in a covariant way in terms of an additional geometric structure. The difference is that we have destroyed much more than in unimodular gravity, and therefore instead of a simple fiducial measure we have to introduce a whole fiducial basis of covariantly constant vectors.

\subsection{Covariant equations of motion}

Let us initially establish that the metric equation \eqref{fqcoincov}, written in the covariant form
\begin{equation}
\label{fqcoveq}
2\nabla_{\alpha}\left( f^{\prime}({\mathfrak Q}) {{\mathfrak P}^{\alpha\mu}}_{\nu}\right) +  f^{\prime}({\mathfrak Q}) \left(Q_{\alpha}{{\mathfrak P}^{\alpha\mu}}_{\nu}+{\mathfrak P}^{\mu\alpha\beta} Q_{\nu\alpha\beta}\right) -  f({\mathfrak Q}) \delta^{\mu}_{\nu}=0,
\end{equation}
remains unaffected by the process of covariantisation. This persistence is a fundamental property observed whenever a tensorial equation has been found in a specific coordinate system.  Therefore, the majority of the derivation steps previously introduced can be replicated with minimal alterations. The key distinction arises from the variation of the metric, yielding $\delta_g Q_{\alpha\mu\nu}= \nabla_{\alpha} \delta g_{\mu\nu}$, with the covariant derivative instead of the partial one. 

The only slightly non-trivial step is the integration by parts producing the new $f^{\prime}({\mathfrak Q}) Q_{\alpha}{{\mathfrak P}^{\alpha\mu}}_{\nu}$ term. Or, in the language of the classical papers \cite{BeltranJimenez:2017tkd, BeltranJimenez:2019tme, BeltranJimenez:2019esp}, we need to {\it define} the action of the covariant derivative onto the tensorial densities such that 
$$\nabla_{\alpha} \sqrt{-g} \equiv \frac12 Q_{\alpha} \sqrt{-g}$$
which also allows us to rewrite the equation (\ref{fqcoveq}) as
\begin{equation}
\frac{2}{\sqrt{-g}}\nabla_{\alpha}\left(\sqrt{-g} f^{\prime}({\mathfrak Q}) {{\mathfrak P}^{\alpha\mu}}_{\nu}\right) +  f^{\prime}({\mathfrak Q}) {\mathfrak P}^{\mu\alpha\beta} Q_{\nu\alpha\beta} -  f({\mathfrak Q}) \delta^{\mu}_{\nu}=0.
\end{equation}

In general, this choice of the rule of differentiating the tensor densities can also be given as an abstract theorem about the integration by parts, see also the Ref. \cite{GSVnewGR} for the purely metric teleparallel case:

\vspace{8pt}

{\bf Theorem.} Given a connection $\Gamma^{\alpha}{}_{\mu\nu} = \mathring{\Gamma}^{\alpha}{}_{\mu\nu} + \delta \Gamma^{\alpha}{}_{\mu\nu}$, in order for integration by parts with the measure of $\sqrt{-g}$ be possible in the most naive way, one needs to define the following action of the covariant derivative onto the tensor densities:
$$\nabla_{\alpha} \sqrt{-g} = - \sqrt{-g}\cdot \delta \Gamma^{\beta}{}_{\beta\alpha}= \sqrt{-g}\cdot \left(\frac12 Q_{\alpha} - T_{\alpha}\right)$$
where $Q_{\alpha}$ and $T_{\alpha}$ are the non-metricity and torsion vectors respectively.

{\bf Proof.} It is evident that the only non-trivial question is about the only tensorial index which is contracted with the index of the derivative. Taking the partial differentiation of the $\sqrt{-g}$ measure produces a $\mathring{\Gamma}^{\beta}{}_{\beta\alpha}$ term, while what we need is $\Gamma^{\beta}{}_{\beta\alpha}$. The difference is precisely what we have to add compared to the case if our derivative was just the Levi-Civita. $\square$

Of course, if integration is treated in terms of differential forms, so that $\sqrt{-g}$ is nothing but components of a volume form, then this result follows from the purely tensorial covariance and the Stokes theorem. However, in case of viewing it as just an absolutely continuous measure with the smooth density $\sqrt{-g}$, we need to give some definition as above, and this desired simple structure of integration by parts makes the purely measure-theoretic definition coincide with the differential forms approach.

\vspace{8pt}

Upon considering the metric equation in both versions of the model, an additional equation emerges within the framework of the covariant theory, namely the connection equation. Naively, this equation appears as a genuinely novel addition resulting from the variation
$$\delta_{\xi }Q_{\alpha\mu\nu} = - g_{\mu\beta} \delta \Gamma^{\beta}{}_{\alpha\nu} - g_{\nu\beta} \delta \Gamma^{\beta}{}_{\alpha\mu}$$
while preserving all other occurrences of the metric tensor. In particular, around the coincident gauge, this equation effectively corresponds to variations solely in $\partial g$, excluding modifications to the metric $g$ itself. The derivation of this equation poses no substantial difficulty.

To start with, in the coincident gauge of the background $\xi^n= x^n$, we have from the equation (\ref{theconn}) the linear variation $\delta \Gamma^{\alpha}{}_{\mu\nu} = \frac{\partial^2 }{\partial x^{\mu} \partial x^{\nu}} \delta\xi^{\alpha} $, and the $\xi$-variation of the action (\ref{genact}) gives the new equation of the form 
\begin{equation}
\label{conneqcoin}
\partial_{\alpha}\partial_{\mu} \left(\sqrt{-g} f^{\prime}\cdot {\mathfrak P}^{\alpha\mu}{}_{\nu}\right)=0.
\end{equation}
It is of the third derivative order. However, the new information it might give can obviously be represented in the second-derivative-order form, by simply subtracting the divergence of the metric equations from it. Less immediately obvious is that, as we will show soon, this difference is an identical zero.

As before, the equation (\ref{conneqcoin}) can also be rewritten in a covariant form as
\begin{equation}
\label{conneqcov}
\left(\nabla_{\alpha} + \frac12 Q_{\alpha}\right)\left(\nabla_{\mu} + \frac12 Q_{\mu}\right) \left( f^{\prime}\cdot {\mathfrak P}^{\alpha\mu}{}_{\nu}\right)=0
\end{equation}
employing up to third derivatives of $g_{\mu\nu}$ and up to fourth derivatives of $\xi^n$.
When the connection is equal to zero, this equation (\ref{conneqcov}) just coincides with the coincident gauge equation (\ref{conneqcoin}). A tensorial equation having been derived in a particular coordinate system must be valid in general. And it is also quite simple to be checked by using the variation of the connection (\ref{convar}). Indeed, we have 
$$\delta_{\xi} {\mathfrak Q}= {\mathfrak P}^{\alpha\mu\nu}\delta Q_{\alpha\mu\nu} = -2 {\mathfrak P}^{\alpha\mu}{}_{\nu} \delta\Gamma^{\nu}{}_{\alpha\mu} = -2 {\mathfrak P}^{\alpha\mu}{}_{\nu}  \nabla_{\alpha} (e^{\nu}_n  \delta e^n_{\mu})$$
with $\delta e^n_{\mu}=\partial_{\mu} \delta \xi^n= \nabla_{\mu} \delta\xi^{n}$, for $\xi^n (x)$ being treated as a collection of scalars, and $\nabla_{\mu} e^{\nu}_n =0$. Then the integration by parts deriving the connection equation goes precisely the same way as around the coincident gauge and ends up with the equation (\ref{conneqcov}).

\subsection{Some technical remarks}
\label{sec:technical}

In this subsection, we would like to discuss the role of Lagrange multipliers in the metric-affine approach to the symmetric teleparallel models, as well as the structure of primary constraints. The latter are only the first step of any Hamiltonian analysis which is much more complicated in full and goes beyond the scope of this paper.

Recall that the teleparallel connection enters the action through the non-metricity tensor components only, and therefore its equation of motion is derived via
$$\delta {\mathfrak Q}=-2 {\mathfrak P}^{\alpha\mu}{}_{\nu} \cdot \delta\Gamma^{\nu}{}_{\alpha\mu}.$$
Recall also that, in the coincident gauge of the usual covariant approach, we have 
$$\delta \Gamma^{\alpha}{}_{\mu\nu} = \frac{\partial^2 }{\partial x^{\mu} \partial x^{\nu}} \delta\xi^{\alpha}$$ 
which produces the familiar equation (\ref{conneqcoin})
$$\partial_{\alpha}\partial_{\mu} \left(\sqrt{-g} f^{\prime}\cdot {\mathfrak P}^{\alpha\mu}{}_{\nu}\right)=0.$$

Since the metric velocity and the connection components come about inside one and the same structure $Q_{\alpha\mu\nu}$, the corresponding momenta immediately satisfy some primary constraints, precisely as in the case of metric teleparallel models \cite{Golovnev:2021omn}. The complication of symmetric teleparallel theories is in their higher-derivative nature. It means that one needs to either use the formal procedure traditionally attributed to Ostrogradsky, or equivalently to perform a rewriting of the Lagrangian in terms of a new variable $v^{n}=\dot{\xi}^n$ such that $${\mathcal L}(\xi^n, \dot\xi^n, \ddot\xi^n)\longrightarrow {\mathcal L}(\xi^n, \dot\xi^n, \dot v^n) + \lambda_n (\dot\xi^n - v^n).$$

In particular, the usual Ostrogradsky procedure for higher derivative Lagrangians requires taking the two different momenta of $\xi^n$:
\be p_n=\frac{\partial\mathcal L}{\partial\dot\xi^n} - \frac{\partial}{\partial t} \left( \frac{\partial\mathcal L}{\partial\ddot\xi^n}\right), \qquad {\mathfrak p}_n=\frac{\partial\mathcal L}{\partial\ddot\xi^n}.
\ee 
Then, the second equation implies a primary constraint. Indeed, the second order time derivatives of $\xi^{n}$ are only inside the following component of the nonmetricity tensor
\be 
Q_{00\alpha}=Q_{0\alpha 0}={\dot g}_{0\alpha} -  g_{\rho\alpha}\cdot {\ddot\xi}^{\rho} -  g_{0\rho} \cdot \partial_{\alpha} \dot\xi^{\rho},
\ee 
therefore we get 
\be
{\mathfrak p}_n =-  \frac{\partial\mathcal L}{\partial Q_{00\alpha}} g_{n \alpha} = - g_{n \alpha} \pi^{0\alpha}
\ee
where $\pi^{0\alpha}$ are the momenta of $g_{0\alpha}$. Note that here we have abandoned the approach of ADM variables. Another important remark is that we have fully imposed identifications of $g_{0\alpha}\equiv g_{\alpha 0}$ and $\pi^{0\alpha}\equiv \pi^{\alpha 0}$, and $Q_{00\alpha}\equiv Q_{0\alpha 0}$, which is not convenient when doing tensorial calculations. However, we need to expose the constraint here only as a matter of principle.

In the Lagrange multiplier approach of putting $v^n={\dot\xi}^n$, we get ${\mathfrak p}_n$ as the momentum of $v^n$, while the $\xi^n$ has then the momentum $\frac{\partial\mathcal L}{\partial\dot\xi^n}+\lambda_n$ which upon substituting the Hamiltonian equation of ${\dot{\mathfrak p}}_n = -\lambda$ turns into the momentum $p_n$ of above.

The need of higher derivatives can be avoided also in terms of our tetrad (\ref{ourconn}). However, then a Lagrange multiplier term $\lambda^{\hphantom{n}\mu\nu}_n \left(\partial_{\mu} e^n_{\nu} - \partial_{\nu} e^n_{\mu}\right) = \lambda^{\hphantom{\alpha}\mu\nu}_{\alpha}\left(\Gamma^{\alpha}{}_{\mu\nu}-\Gamma^{\alpha}{}_{\nu\mu}\right)$ has to be added to the action in order to set the torsion to zero, or to impose the form (\ref{basicforms}) of the tetrad. Using the variation (\ref{convar}), it yields an equation for the tetrad,
$$2\partial_{\alpha} \left(\sqrt{-g} f^{\prime}\cdot {\mathfrak P}^{\alpha\mu}{}_{\nu}\right) - \partial_{\alpha} \left(\sqrt{-g}(\lambda_{\nu}^{\hphantom{\nu}\alpha\mu} - \lambda_{\nu}^{\hphantom{\nu}\mu\alpha})\right)=0.$$
Obviously, taking a divergence of this equation immediately produces the one which we had before (\ref{conneqcoin}). And vice versa, the equation (\ref{conneqcoin}) states that the rank-2 tensor $2\partial_{\alpha} \left(\sqrt{-g} f^{\prime}\cdot {\mathfrak P}^{\alpha\mu}{}_{\nu}\right)$ had zero divergence. At least locally, it means that it can be represented in the shape of $\partial_{\alpha} \left(\sqrt{-g}(\lambda_{\nu}^{\hphantom{\nu}\alpha\mu} - \lambda_{\nu}^{\hphantom{\nu}\mu\alpha})\right)$, similar to a divergenceless vector being a curl of another one. 

This is an example of how the Lagrange multiplier approach works. Note also that this is a case when the derivatives of Lagrange multipliers give us precisely the physics we want. In this sense the claims of the recent paper \cite{Lavinia} that terms of this kind invalidate the Hamiltonian analysis seem unsubstantiated. Although we must always give accurate account of the counting of the degrees of freedom, the terms with derivatives on Lagrange multipliers can have a physical meaning behind.

The realisation that the connection can be expressed in the way (\ref{ourconn}) of a teleparallel tetrad also makes it possible to present the covariant Hamiltonian analysis of symmetric teleparallel theories in a nice form. It follows that the canonical momenta are
\begin{align}
    \pi^{\mu\nu}:=\frac{\partial L}{\dot{g}_{\mu\nu}}=\frac{\partial L}{\partial Q_{0\mu\nu}}
\end{align}
and
\begin{align}
    \Pi_n{}^\rho:=\frac{\partial L}{\partial \dot{e}^n{}_\rho}=\frac{\partial L}{\partial Q_{0\mu\nu}}\frac{\partial Q_{0\mu\nu}}{\partial \dot{e}^n{}_\rho} + \sqrt{-g}(\lambda_n^{\hphantom{n}0\rho}- \lambda_n^{\hphantom{n}\rho 0} )=-2\pi^{\mu\nu}e_n{}^{\lambda}\delta^\rho_{(\mu}g_{\nu)\lambda} +2\sqrt{-g}\cdot\lambda_n^{\hphantom{n}[0\rho]}
    \end{align}
with the usual symbols of symmetrisation and antisymmetrisation, and with the usual way of treating the $g_{\mu\nu}$ and $g_{\nu\mu}$ components separately.
    This introduces the primary constraints
    \begin{align}
    \label{eq:CovPC}
        C_n{}^\rho=\Pi_n{}^\rho+2\pi^{\mu\nu}e_n{}^\lambda \delta^\rho_{(\mu}g_{\nu)\lambda} - 2\sqrt{-g}\cdot\lambda_n^{\hphantom{n}[0\rho]}\approx 0,
    \end{align}
on top of the obvious zero canonical momenta of the Lagrange multipliers themselves.

In what concerns the full-fledged metric-affine approach, fixing the torsion to zero by a Lagrange multiplier is not necessary at all. We can just treat the connection as  symmetric in the lower indices $\Gamma^{\alpha}{}_{\mu\nu}=\Gamma^{\alpha}{}_{\nu\mu}$, similar to how we always deal with the metric. At the same time, for putting the curvature to zero, we add a new term, $\lambda_{\alpha}^{\hphantom{\alpha} \beta\mu\nu} R^{\alpha}_{\hphantom{\alpha} \beta\mu\nu}$ with $\lambda_{\alpha}^{\hphantom{\alpha} \beta\mu\nu}=-\lambda_{\alpha}^{\hphantom{\alpha} \beta\nu\mu}$. Around the coincident gauge, and remembering that the connection variation is symmetric, we get
$$2\sqrt{-g} f^{\prime}\cdot {\mathfrak P}^{\alpha\mu}{}_{\nu} + \partial_{\rho} \left(\sqrt{-g}(\lambda_{\nu}^{\hphantom{\rho}\alpha\rho\mu} + \lambda_{\nu}^{\hphantom{\rho}\mu\rho\alpha})\right)=0.$$
We immediately see that the shape of the second term is precisely requiring that the double divergence of the first term vanishes, i.e. the equation (\ref{conneqcoin}) we know. At the same time, the primary constraints express the connection momenta in terms of the Lagrange multipliers.

\subsection{The covariant "conservation" laws}

Unlike the covariantisation process observed in metric teleparallelism, the $\xi$-variation naively looks like a  completely new procedure. It induces a transformation which changes $Q_{\alpha\mu\nu}$ without changing the metric itself. However, it is a common non-trivial fact about the Noether identities for diffeomorphisms. Normally, the Lagrangian density and the measure have non-vanishing changes as explicit functions of coordinates, while the action integral stays invariant; a parallel that can be applied to the current context. Let us perform, in a bounded region, a simultaneous change of coordinates in the metric and the $\xi^{\alpha}$ represented by $\delta x^{\mu}=\zeta^{\mu}$, then the change corresponds to
$$\delta g_{\mu\nu} =- \mathring{\nabla}_{\mu} \zeta_{\nu} - \mathring{\nabla}_{\nu} \zeta_{\mu}\qquad \mathrm{and} \qquad \delta \xi^{\alpha}=- (\partial_{\mu} \xi^{\alpha}) \cdot\zeta^{\mu}.$$
With it, the Lagrangian density changes as a full-fledged scalar, and therefore the action is automatically invariant. Subsequently, we immediately find the following identical relation between the equations:
\be
2 \mathring{\nabla}_{\mu} \left( \frac{\delta S}{\delta g_{\mu\nu}}\right) - (\partial^{\nu} \xi^{\alpha}) \cdot \frac{\delta S}{\delta \xi^{\alpha}}=0.
\ee
We see that fulfillment of the connection equation implies that the metric equation is automatically divergenceless. And as long as the matrix of $(\partial_{\mu} \xi^{\alpha})$ is invertible, as it must always be, those two conditions are just equivalent.

Therefore, as long as we are dealing with the equations of motion in vacuum, the connection equation does not give us any new insights. If we incorporate an arbitrary matter source to the right hand side, the connection equation is simply equivalent to insisting on its covariant "conservation". In particular, for a model with the action featuring modified symmetric teleparallel gravity and matter which only interacts with the metric in a covariant way, the covariantisation changes nothing and the connection equation gives no new information. At the same time, this equivalence comes at the price of getting a higher-derivative action.

Note in passing that options of gravity theories with non-conserved matter have received a significant attention in the literature, for example in the shape of Rastall gravity \cite{Rastall}. At the same time, the latter is just a mere superficial modification of gravity, or more precisely, no modification at all \cite{Visser, GaR}, as the equations remain identical while the energy-momentum tensor is redefined by a trace subtraction. However, the non-covariantised modified symmetric teleparallel gravity allows for more meaningful deviations from covariant conservation laws.

\subsection{The case of STEGR}

It is rightful to also explicitly check that in STEGR no equation follows from the variation of $\xi$, as it is expected since this field enters only in the surface term. In the coincident gauge, this equation (\ref{conneqcoin}) would be of the form
$$\partial_{\alpha}\partial_{\mu}\left(\sqrt{-g}P^{\alpha\mu}{}_{\nu}\right)=0$$
with the STEGR non-metricity conjugate
\begin{multline}
P_{\alpha\mu\nu}=\frac12 Q_{\alpha\mu\nu} - \frac12 Q_{\mu\nu\alpha} - \frac12 Q_{\nu\mu\alpha} -\frac12 g_{\mu\nu} Q_{\alpha} + \frac12 g_{\mu\nu} {\tilde Q}_{\alpha} + \frac14 g_{\alpha\mu} Q_{\nu} + \frac14 g_{\alpha\nu} Q_{\mu}\\
=-\frac12 Q_{\nu\mu\alpha} - \frac14 g_{\mu\nu}  Q_{\alpha} + \frac12 g_{\mu\nu} {\tilde Q}_{\alpha} + \frac14 g_{\alpha\mu} Q_{\nu} +\quad \mathrm{antisymmetric\quad in}\quad \alpha\leftrightarrow\mu\quad \mathrm{part}.
\end{multline}
Using basic formulae such as $\quad Q_{\mu}=2\frac{\partial_{\mu} \sqrt{-g}}{\sqrt{-g}},\quad $ $Q_{\alpha}^{\hphantom{\alpha}\mu\nu}=-\partial_{\alpha}g^{\mu\nu},\quad$ $Q_{\mu}^{\hphantom{\mu}\alpha\mu}={\tilde Q}^{\alpha},\quad $ we get
\begin{multline*}
\sqrt{-g}P^{(\alpha\mu)}{}_{\nu}= \sqrt{-g}\left( -\frac12 Q_{\nu}^{\hphantom{\nu}\mu\alpha} - \frac14 \delta^{(\mu}_{\nu}  Q^{\alpha)} + \frac12 \delta^{(\mu}_{\nu} {\tilde Q}^{\alpha)} + \frac14 g^{\alpha\mu} Q_{\nu}\right)\\
= \frac12 \left(\vphantom{\frac14} \sqrt{-g} \partial_{\nu} g^{\alpha\mu} -  \delta^{(\mu}_{\nu}  g^{\alpha)\beta}\partial_{\beta} {\sqrt{-g}} -\sqrt{-g}  \delta^{(\mu}_{\nu} \partial_{\beta} g^{\alpha)\beta} + g^{\alpha\mu} \partial_{\nu} {\sqrt{-g}}\right) 
\end{multline*}
from where it is easy to see that $\partial^2_{\alpha\mu}\left(\sqrt{-g}P^{\alpha\mu}{}_{\nu}\right)\equiv 0$, with no extra tricks, by merely expanding the whole expression. For example, one can collect the terms with 0, 1, 2, and 3 derivatives acting on $\sqrt{-g}$ separately, and see that they all cancel each other.

\section{A toy model example}
\label{sec:toy}

Let us give an illustration of how the equations of motion work. A simple model would be the one with only $c_1$ non-zero. In the coincident gauge it has the following action
\be
S=\int d^4 x \sqrt{-g}\cdot  Q_{\alpha\mu\nu}Q^{\alpha\mu\nu} = \int d^4 x \sqrt{-g}\cdot  (\partial_{\alpha_1}g_{\mu_1\nu_1})(\partial_{\alpha_2}g_{\mu_2\nu_2}) g^{\alpha_1 \alpha_2} g^{\mu_1 \mu_2} g^{\nu_1 \nu_2}
\ee
with the evident equations of motion
\be 
4\partial_{\alpha} \left(\sqrt{-g} Q^{\alpha\mu\nu}\right) + \left(2Q^{\mu\alpha\beta}Q^{\nu}_{\hphantom{\nu}\alpha\beta}+ 4Q^{\alpha\mu\beta}Q^{\hphantom{\alpha}\nu}_{\alpha\hphantom{\nu}\beta}- Q^{\alpha\beta\rho}Q_{\alpha\beta\rho} g^{\mu\nu}\right) =0.
\ee
We will take them in a different form, with a mixed position of indices as in Eq. \eqref{fqcoin}:
\be
\frac{4}{\sqrt{-g}}\partial_{\alpha} \left(\sqrt{-g} Q^{\alpha\mu}{}_{\nu}\right) +\sqrt{-g} \left(2Q^{\mu\alpha\beta}Q_{\nu\alpha\beta}- Q^{\alpha\beta\rho}Q_{\alpha\beta\rho} \delta^{\mu}_{\nu}\right) =0.
\ee 
This model is not physically viable for sure, since it has ghosts. Due to the metric not being positive definite, its mixed components appear to have the sign of kinetic energy opposite to that of the temporal and purely spatial ones. Therefore, we will use it only for illustrative purposes. Leaving aside the issue of ghosts, it is otherwise perfectly healthy. As long as the metric is non-degenerate, it does not have any constraints at all, and therefore no changes in the symplectic structure.

\subsection{Vacuum cosmology}

Now we will investigate the behaviour of the standard spatially-flat cosmology in vacuum in our previous toy model. Substituting the metric ansatz
\be 
g_{\mu\nu}=\mathrm{diag} (N^2(t), -a^2(t), -a^2(t), -a^2(t) )
\ee
we easily get the following equations (after having multiplied them by $\frac{N^2}{4}$):
\be 
2\frac{\ddot N}{N} - 3\frac{{\dot N}^2}{N^2} + 6\frac{{\dot a}\dot N}{aN} + 3\frac{{\dot a}^2}{a^2}=0,
\label{c1cosmo1}
\ee
\be 
2\frac{\ddot a}{a} + \frac{{\dot a}^2}{a^2} - 2\frac{{\dot a}\dot N}{aN} - \frac{{\dot N}^2}{N^2}=0.
\label{c1cosmo2}
\ee
To start with, we notice the absence of time-reparametrisation invariance. An attempt to establish it in the physical time, $N=1$, immediately leads to Minkowski spacetime in the first equation, as in the usual cosmology. And an attempt of using the conformal time, $a=N$, leads to the same conclusion, now by combining the two equations. Nevertheless, there are definitely some other solutions of this system, as we prove in the following.

In effect, there is a non-trivial and simple vacuum solution. If we assume that $a=N^k$, it is a simple task to find a cubic equation for $k$ with the only real root of $k=-1$. Substituting the ansatz $$N=\frac{1}{a},$$ 
the equations \eqref{c1cosmo1} and \eqref{c1cosmo2} transform into $-2\left(\frac{\ddot a}{a} + \frac{{\dot a}^2}{a^2}\right)=0 $ and $2\left(\frac{\ddot a}{a} + \frac{{\dot a}^2}{a^2}\right)=0 $, respectively. We then have a solution of $a\propto\sqrt{t}$ with the metric 
$$g_{\mu\nu} dx^{\mu} dx^{\nu} = \frac{1}{t}\cdot dt^2 - t\cdot d\vec{x}^2$$
which is a constant-rate expansion $a\propto T$ in the physical time $T\propto \sqrt{t}$.

Note that, for the solution we have just found, the connection equation (\ref{conneqcoin})
$$4\partial_{\alpha}\partial_{\mu} \left( \sqrt{-g} Q^{\alpha\mu}{}_{\nu}\right)=0$$
is also satisfied. Indeed, the equation takes the following form
\begin{equation}
\label{toyconn}
\partial^2_0 \left(\frac{a^3 \dot N}{N^2}\right)=\partial_0^2 \left(a^3 {\dot a}\right)=0
\end{equation}
which, given $a\propto\sqrt{t}$, states that the second derivative of a linear function is actually zero. 

\subsection{Cosmology with matter}

With matter, we need to put $N^2 T^{\mu}_{\nu}$   to the right hand side of the equations above (a constant pre-factor, in particular with the gravitational constant, isn't of any importance for us now). Assuming again that
$$N=\frac{1}{a}$$
we get
$$\left(\frac{\ddot a}{a} + \frac{{\dot a}^2}{a^2}\right)=\frac{\rho}{a^2},$$
$$-\left(\frac{\ddot a}{a} + \frac{{\dot a}^2}{a^2}\right)=-\frac{p}{a^2}.$$

We see that this Ansatz allows only for superstiff matter with an equation of state $p=\rho$. In this case, if we want to impose covariant energy conservation, it means to set the energy density to
$$\rho\propto\frac{1}{a^6}.$$ 
When we substitute this behaviour of $p=\rho$ to the right hand sides, $$
\left(\frac{\ddot a}{a} + \frac{{\dot a}^2}{a^2}\right)\propto\frac{1}{a^8},
$$ there is a straightforward solution of $a\propto t^{1/4}$. Meanwhile, the information encoded in the connection equation \eqref{toyconn}, in the case of $N=\frac{1}{a}$, simply states that $\partial^3_0 a^4 = 0$, which is evidently true.

Moreover, note that the equation 
$$a {\ddot a} + {\dot a}^2 \propto \frac{1}{a^6},$$ 
or the cosmology with $N=\frac{1}{a}$ and superstiff matter of $\rho=p\propto\frac{1}{a^6}$, looks like requiring quite some (mild) effort to be fully solved. However, the connection equation (\ref{toyconn}) in this case has the obvious solution of 
$$a(t) = \left(c_1 t^2 + c_2 t + c_3\right)^{\frac14}.$$ 
And we immediately see that this is precisely the solution for the metric equation since, upon substitution, we get $$a {\ddot a} + {\dot a}^2=\frac{4c_1 c_3 - c_2^2}{8a^6},$$ 
reproducing also the $a(t)\propto\sqrt{t-t_0}$ vacuum solution when $c_2=\pm 2\sqrt{c_1 c_3}$.

All in all, we confirm that the covariantisation procedure does not introduce any substantial changes, given that the matter content satisfies the covariant conservation laws. The covariant version of symmetric teleparallel gravity simply prohibits the inclusion of a source with $\mathring{\nabla}_{\mu} T^{\mu\nu}\neq 0$ to the right hand side. On the other hand, similar to the potential benefits of a non-zero spin connection in metric teleparallel gravity \cite{LorGov}, the covariant version can also be technically advantageous, even in absence of non-trivial $\xi^{\mu}$. In particular, in the example above, solving the connection equation was much easier than working directly with the metric equation, despite being fully equivalent in the end.

\section{Comments on the case of $f(\mathbb Q)$}
\label{sec:fq}

Precisely as in the case of $f(\mathbb T)$ theory, the simplest non-linear generalisations of STEGR do have an accidental gauge symmetry in the weak gravity limit. Indeed, for fluctuations around $g_{\mu\nu}=\eta_{\mu\nu}$ in the coincident gauge, the non-zero non-metricity tensor appears only in perturbations. Therefore, the $\mathbb Q$ scalar is quadratic in perturbations, and we must take a linear approximation to the function $f(\mathbb Q)$ in the quadratic action thus coming back to STEGR or just GR. In other words, the linear weak gravitational waves in $f(\mathbb Q)$ are no different from the case of GR. As we see, this is immediately obvious and makes detailed investigations of these matters \cite{Salva} look rather strange.

Beyond the weak gravity limit, let us start with the standard cosmology \cite{BeltranJimenez:2019tme} in $f(\mathbb Q)$, which is a very peculiar case. We can take the metric
\be 
ds^2 = N^2 (t) dt^2 - a^2 (t) (dx^2+dy^2+dz^2)
\ee  
in the coincident gauge. The only non-zero components of the non-metricity and the superpotential are then $Q_{000}=2N\dot N$, $Q_{0ij}=-2a{\dot a}\delta_{ij}$, $Q_0 = 2\frac{\dot N}{N} + 6\frac{\dot a}{a}$, ${\tilde Q}_0=2\frac{\dot N}{N}$, $P_{0ij}=2a{\dot a}\delta_{ij}$, $P_{i0j}=P_{ij0}=-a^2 \left(\frac{\dot N}{N} + 2\frac{\dot a}{a}\right)\delta_{ij}$. In particular, the non-metricity scalar
$${\mathbb Q}=-6\frac{{\dot a}^2}{a^2 N^2}$$
appears to be a genuine, time-reparametrisation-invariant, scalar.

The temporal equation yields the condition $f=2{\mathbb Q} f^{\prime}$ in  vacuum. Except when $f\propto \sqrt{-\mathbb Q}$, this condition requires that $\mathbb Q$ is constant . Generically, if $f(0)=0$ then Minkowski is a solution with $\mathbb Q=0$, though other constant $\mathbb Q$ configurations are possible also for different functions, resulting in a de Sitter geometry. Remarkably, it is observed that  $\mathbb Q$ turns out to be time-reparametrisation invariant, and is constant for de Sitter. Consequently, these solutions correspond to pure GR solutions with an effective cosmological constant.

Considering the inclusion of matter, a very interesting aspect is that its covariant conservation is automatically required in this case. Since only the time-dependence is there, we see that the connection equation takes the form of 
\be 
\partial^2_0 \left(f^{\prime}\sqrt{-g}{P^{00}}_{\mu}\right)=0.
\ee 
This equation is trivially satisfied because $P_{00\mu}=0$ identically. A fundamental reason of this result is the following: due to rotational symmetry, we cannot have non-zero $P_{00i}$, while $P_{000}=0$ is akin to having the lapse non-dynamical in GR.

Actually, the fulfillment of the connection equation as an identity, or identical divergencelessness of the metric equation, is a more general statement. It is analogous to the  statement in $f(\mathbb T)$ gravity which states that any diagonal tetrad, whose components depend on only one of the coordinates in which it is diagonal, automatically solves the antisymmetric part of the equations of motion \cite{CairoBH}. We can formulate a similar property as an additional theorem
\vspace{8pt}

{\bf Theorem}. The $f(\mathbb Q)$ connection equation (and the Levi-Civita divergence of the metric equation) gets automatically satisfied upon substitution of a metric Ansatz which, in the coincident gauge, is diagonal and with the components depending on only one of the chosen coordinates.

{\bf Proof}. In effect, we have only diagonal metric components $g_{\mu\mu}={\mathcal F}_{\mu}(x^{\chi})$ (no summation) where $\chi$ is the index of the only coordinate on which the metric depends. Then the equation $\partial^2_{\chi\chi} \left(\sqrt{-g}f^{\prime}P^{\chi\chi}{}_{\mu}\right)=0$ is automatically satisfied simply because $P_{\chi\chi\mu}\equiv 0$ for every index $\mu$ including $\chi$ itself, as can be checked by direct inspection. $\square$

\vspace{8pt}

Due to this reason, the role of the connection equation can not be illustrated in simple $f(\mathbb Q)$ cosmology, unlike our toy model example above. However, it would be instructive to explicitly see how it works in general. Such a statement has already been made in Ref. \cite{De:2022jvo}, although in a much less transparent way.

Let's consider the metric equation of motion in the geometric form given in Eq. \eqref{eqfq}. Then its divergence reads
$$f^{\prime\prime} \mathring{G}^{\mu}_{\nu}\partial_{\mu} {\mathbb Q}-\frac12 f^{\prime\prime} {\mathbb Q} \partial_{\nu} {\mathbb Q} - {\mathring{\nabla}}_{\mu} \left(f^{\prime\prime}P^{\alpha\mu}{}_{\nu}\partial_{\alpha}\mathbb Q\right).$$
By using an expression for the Einstein tensor, see the Eq.  (\ref{Ricciten}),
$$\mathring{G}^{\mu}_{\nu}= - \frac{1}{\sqrt{-g}}\partial_{\alpha}\left(\sqrt{-g} P^{\alpha\mu}{}_{\nu}\right) - \frac12 P^{\mu\alpha\beta} Q_{\nu\alpha\beta} + \frac12 {\mathbb Q}\delta^{\mu}_{\nu}$$
and a simple observation (straightforwardly transforming the term with $\mathring{\Gamma}^{\beta}_{\mu\nu}$) of
$${\mathring{\nabla}}_{\mu} \left(f^{\prime\prime}P^{\alpha\mu}{}_{\nu}\partial_{\alpha}\mathbb Q\right) = \frac{1}{\sqrt{-g}} \partial_{\mu} \left(\sqrt{-g} f^{\prime\prime}P^{\alpha\mu}{}_{\nu}\partial_{\alpha}\mathbb Q\right) - \frac12 f^{\prime\prime} Q_{\nu\mu\beta} P^{\alpha\mu\beta} \partial_{\alpha} \mathbb Q,$$
we obtain the equations' divergence as
$$-f^{\prime\prime}\left(\frac{\partial_{\mu} {\mathbb Q}}{\sqrt{-g}}\partial_{\alpha}\left(\sqrt{-g} P^{\alpha\mu}{}_{\nu}\right) + \frac{\partial_{\alpha} {\mathbb Q}}{\sqrt{-g}}\partial_{\mu}\left(\sqrt{-g} P^{\alpha\mu}{}_{\nu}\right) \right) - f^{\prime\prime\prime} P^{\alpha\mu}{}_{\nu} (\partial_{\alpha}\mathbb Q) (\partial_{\mu}\mathbb Q).$$

At the same time, if in the connection equation (\ref{conneqcoin})
$$\frac{1}{\sqrt{-g}}\partial_{\alpha}\partial_{\mu}\left(\sqrt{-g} f^{\prime} P^{\alpha\mu}{}_{\nu}\right)=0$$
we take into account that for the STEGR superpotential $\partial_{\alpha}\partial_{\mu}\left(\sqrt{-g}  P^{\alpha\mu}{}_{\nu}\right)\equiv 0$, it also acquires (minus) the same shape.

As a relatively simple example of this statement in $f(\mathbb Q)$ theories, one can take a look at the unusual cosmologies which have been studied recently \cite{Dimakis:2022rkd, Paliathanasis:2023nkb}. To the best of our knowledge, their explicit coincident gauge form isn't presented anywhere, except for the first family of connections which corresponds to the standard cosmology we mentioned above. However, without discussing how cosmological the other two families of solutions really are, let us have a look at the ones which correspond to their second family of connections. Its equations of motion are Eqs. (11 - 13) in Ref. \cite{Paliathanasis:2023nkb}. One can do the following steps. 1. Differentiate their Eq. (11) with respect to time. 2. Find $\dot H$ from their Eq. (12) and substitute it into the time derivative obtained. 3. In place of the non-differentiated entrance of the non-metricity scalar, substitute its expression from just above the formulae. 4. Note that what it yields is precisely $\frac32 \gamma$ times their Eq. (13). Contemplating this procedure a bit, one can see that the connection equation was indeed obtained by taking a Levi-Civita divergence of the metric equation.

\section{Discussion}
\label{sec:dis}

An important concern that the implementation of the additional variable $\xi^{\mu}$ raises is that its appearance in the Lagrangian for any modified symmetric teleparallel gravity goes with second order derivatives. Although for STEGR this is harmless, in the case of modified symmetric teleparallel gravities such terms give fourth order equations of motion for the $\xi$ field, and third order for the metric. We would like to discuss some theoretical issues that have not been posed in the previous literature.\\ 

\textbf{Degrees of freedom.} In order to have a rough sketch  on the counting of degrees of freedom for the geometric trinity of gravity, we refer the reader to Figure 2 in \cite{BeltranJimenez:2019esp}. Here we would like to improve the naive counting of degrees of freedom presented there from the point of view of constrained Hamiltonian systems. As it was presented elsewhere \cite{Golovnev:2021omn}, in the covariant version of TEGR there are $16$ independent variables from the tetrad, and additional $6$ components of the Lorentz matrices $\Lambda^{a}{}_{b}$ that introduce the so-called "inertial" connection. Their appearance introduces $6$ primary first-class constraints $C^{\prime}_{ab}$ in the covariant version of the model, while in both approaches there are $6$ other constraints, $C_{ab}$ or $C^{\text{cov}}_{ab}$, coming from the pseudo Lorentz invariance of the TEGR Lagrangian with respect to pure tetrad rotations \cite{Golovnev:2021omn}. They have incidentally been called, respectively, Lorentz transformations of type I and II \cite{Blixt:2022rpl}. In addition to them, $4$ primary constraints $\Pi^{0}_a$ coming from the absence of time derivatives of the $e^{a}{}_0$ components of the tetrad (equivalent to the nondynamical character of the lapse and shift functions) produce $4$ secondary constraints, the well-known Hamiltonian and momenta constraints $C_0$ and $C_i$. Since all constraints are first class, the counting of degrees of freedom in TEGR goes as follows. The number of pairs of canonical variables is $16+6=22$, and the number of first class constraints is $20$. Therefore, we obtain $22-20=2$ propagating degrees of freedom.

The STEGR Lagrangian written in an arbitrary gauge has $10 + 4$ independent components coming from $g_{\mu\nu}$ and $\xi^{\mu}$. The theory has $4$ primary constraints associated with the absence of dynamics for lapse and shift in the Lagrangian, which at the same time generate $4$ secondary constraints reflected in one Hamiltonian $C^{0}$ and three momenta $C^{i}$ constraints. The introduction of $\xi^{\mu}$ would generate four additional first-class constraints, if this variable appeared with first order derivatives in the Lagrangian. However, since it actually comes with second order derivatives, it is necessary to either introduce the Ostrogradsky procedure, or to formulate the problem in terms of  Lagrange multipliers and auxiliary tetrad fields as discussed in Section \ref{sec:technical}. We foresee that eight primary (first-class) constraints would appear from such a procedure, four for the components $\xi^{\mu}$ and four for its time derivatives $\dot{\xi}^{\mu}$. Then the counting goes as follows: from the $10+4+4=18$ pairs of canonical variables we must remove $4+4$ constraints from lapse and shift non-dynamical behaviour and consequent secondary constraints, plus $4+4$ constraints associated with the gauge symmetries produced by the variables $\xi^{\mu}$ and $\dot{\xi}^{\mu}$ inducing the covariantisation. Therefore, the physical number of dof is $18-4\times 4=2$, the same as in GR and TEGR.

Summarising it once more, the covariant procedure for simultaneous diffeomorphisms introduces four new canonical fields which in theory should introduce four new primary constraints. The nature of these constraints is the same as for metric teleparallel gravities: they represent the freedom in the connection that is removed when going to the coincident gauge. The problem, as it might already be clear, is that the $\xi^{\mu}$ is introduced with second order derivatives in the action.  As shown  explicitly in Section \ref{sec:technical}, for counting of degrees of freedom in the nonlinear case, four new primary constraints can be found (see Equation \eqref{eq:CovPC}) at the cost of introducing Lagrange multipliers. Alternatively, one can rely on  the Ostrogradsky procedure. For both options the Hamiltonian analysis will be more intricate compared to the metric teleparallel case.  \\

\textbf{Equivalence classes of solutions.} In metric teleparallel gravity we can define an \textit{equivalence class of solutions of the covariant theory} in the following way. Consider the field equations of a metric teleparallel theory in the Weitzenböck gauge with the solution $e^a{}_\mu$. Then the couple $\{e^b{}_\mu \Lambda_b{}^a, \omega^a{}_{b\mu} \}$ where $\omega^a{}_{b\mu}$ is defined with the same Lorentz matrix $\Lambda$ is considered as part of the same equivalence class of solutions. Of course, this is nothing but just calling configurations obtained by a gauge symmetry transformation equivalent.  For trivial $\Lambda$ we have the solution $\{e^b{}_\mu,0\}$ as one solution in this equivalence class, and the findings of \cite{Golovnev:2023yla, Blixt:2022rpl} show that a solution of the form $\{e^b{}_\mu \Lambda_b{}^a, \omega^a{}_{b\mu} \}$ is basically the same solution after a change of variables. Of course, in TEGR even isolated transformations of the metric only, or of the spin connection only, still don't influence the equations at all.

Note that in modified metric teleparallel theories, like $f(\mathbb T)$ or NGR,  we can sometimes find that both $\{e^b{}_\mu , 0 \}$ and $\{e^b{}_\mu L_b{}^a, 0 \}$ are solutions for some specific tetrad $e$ and Lorentz matrix $L$. They are, thus, two different solutions belonging to different equivalent classes of solutions, $\{e^b{}_\mu \Lambda_b{}^a, \omega^a{}_{b\mu} \}$ and $\{e^c{}_\mu L_c{}^b \Lambda_b{}^a, \omega^a{}_{b\mu} \}$ respectively. This led to the interest in study of so-called ``non-trivial Minkowski solutions'' in \cite{Golovnev:2020nln}. To some extent, these non-equivalent solutions with the same metric are related with the so-called remnant symmetries, which are an interesting topic which goes beyond this paper. 
 
In symmetric teleparallel gravity the mapping of solutions goes as follows. We consider the field equations of a symmetric teleparallel theory in the coincident gauge where, for a particular choice of coordinates $\xi^{\mu}=x^{\mu}$ the affine connection vanishes. A solution of the equations of motion is then represented by the couple $\{g_{\mu\nu},0\}$, since the non-metricity tensor directly depends on it: $Q_{\alpha\mu\nu}=\partial_{\alpha}g_{\mu\nu}$, while the connection just vanishes. A change of coordinates applied simultaneously to the metric and the affine connection modifies both, and now the same solution is represented by the pair $\{\tilde{g}_{\mu\nu},\Gamma^\rho{}_{\mu\nu}(\xi^{\mu}) \}$. Note that sometimes it might be easier to find a solution to the field equations of the latter form, and in principle one may then always relate this to the coincident gauge case by finding the correct coordinate transformation \cite{Bahamonde:2022zgj}.

Analogous to \cite{Blixt:2022rpl} we can here also talk about diffeomorphisms of type I and type II. Type I, which is always a gauge symmetry of these models, is defined as simultaneous transformation of both $g_{\mu\nu}$ as a tensor and $\xi^{\mu}$ as a collection of scalars. Type II, invariance under which is fully there only in STEGR, is defined as transformations of the tensor $g_{\mu\nu}$ alone. Combining the type I with type II, one can also get transformations of the $\xi^{\mu}$ scalars alone. In modified symmetric teleparallel models, one can then talk about the mystery of remnant symmetries as has already been vividly discussed in the case of metric teleparallel theories \cite{Ferraro:2020tqk}. \\

{\bf General teleparallel theories.} Finally, we would like to mention some thoughts regarding the case of general teleparallel theories written in terms of tetrads. In this case, there also exists a frame for which the spin connection vanishes. It can be obtained simply because, by the very meaning of vanishing curvature, there exists a basis of covariantly constant vectors. However, the affine connection on the manifold cannot be made equal zero since generically there is torsion. In these general teleparallel theories we must treat the tetrad and the metric as two independent variables. If the metric is restricted by the tetrad being pronounced orthonormal (or with some other fixed matrix of its scalar products \cite{Golovnev:2023yla}), then we are back to the metric teleparallel. If the tetrad is restricted to be composed of coordinate vectors (i.e. no anholonomy), then we are back to symmetric teleparallel. Note though that, in the most general case, the discussion of type I and type II symmetries becomes then a bit more involved since we do have non-trivial incarnations of both diffeomorphisms and local linear transformations of tetrads.

Let us also note in passing that, as we saw in Section \ref{sec:technical}, a possible non-Ostrogradskian way to get rid of the higher derivatives in the covariant symmetric teleparallel framework would be to write the action with the metric $g_{\mu\nu}$ and the connection coefficient $\Gamma^{\alpha}_{\mu\nu}=e^{\alpha}_n \partial_{\mu} e^n_{\nu}$ in terms of an arbitrary tetrad, therefore having both the metric and the tetrad as dynamical variables, and then to add a constraint term $\lambda^{\mu\nu}_n \left(\partial_{\mu} e^n_{\nu} - \partial_{\nu} e^n_{\mu}\right)$, which guarantees that the vectors $e^n$ can be written as gradients of some scalars. Of course, the price to pay then is the necessity of working with the Lagrange multipliers, in front of the non-integrable constraint.

\section{Conclusions}
\label{sec:ccl}

In this work, we have provided an overview of symmetric teleparallel gravity models and the covariantisation procedure which turns out to only impose covariant conservation on otherwise arbitrary matter source. The introduction of the set of four scalars $\xi^{\mu}$ for this purpose presents both advantages and theoretical challenges. Regarding the latter, the appearance of the additional fields $\xi^{\mu}$ with second order derivatives in the Lagrangian blurs a clean physical interpretation as in the case of covariance in teleparallel gravity. Nonetheless, it is possible to circumvent this issue by the introduction of (non-orthonormal) tetrads. Namely, any teleparallel geometry can be characterised by a covariantly constant tetrad, or a tetrad with zero spin connection in the fully covariant language.

In other words, we have given a Weitzenb\"ock-like description of a general teleparallel geometry. Of course, it has no curvature. In the symmetric teleparallel case, there is no torsion either, $\partial_{\mu} e^n_{\nu} - \partial_{\nu} e^n_{\mu}=0$, which means that the tetrad takes the form of a gradient: $e^n_{\mu}=\frac{\partial\xi^n}{\partial x^{\mu}}$. Basically, the symmetric teleparallel connection describes the parallel transport of a Minkowski spacetime, with $\xi^n$ being its Cartesian coordinates, and with the teleparallel tetrad constructed from the coordinate vectors of those coordinates. The covariant rewriting of the framework switches it to arbitrary coordinates, while the ``Cartesian" ones become some scalar functions on the manifold.

We have exhibited the behaviour of the equations of motion in the covariant formulation in a FLRW cosmology, for two symmetric teleparallel models: the first as a toy model that considers only one term of the quadratic combination of the non-metricity tensor making up the non-metricity scalar, and the second model consisting of $f(\mathbb Q)$ gravity. In both cases we have given an explicit illustration for our general proof that the only effect of the covariantisation comes in the requirement of covariant conservation for the matter source. Consequently, we raise the pertinent question whether it is worthwhile pursuing the covariant approach. On one hand, there is a demand for non-conserved theories in nowadays research in modified gravity. But on the other hand, this prohibition costs higher derivatives in the action, with all the potential consequences.

\section*{Acknowledgments}
MJG has been supported by the Estonian Research Council grant PSG910.


\begin{thebibliography}{999}

\bibitem{Aldrovandi:2013wha} 
R.~Aldrovandi and J.~G.~Pereira, Teleparallel Gravity : An Introduction, Fundam.\ Theor.\ Phys.\  {\bf 173} (2013).

\bibitem{Golovnev:2018red} 
A.~Golovnev, Introduction to teleparallel gravities, \href{https://arxiv.org/abs/1801.06929}{\textcolor{magenta}{Proceedings of the 9th Mathematical Physics Meeting: School and Conference on Modern Mathematical Physics in Belgrade, September 2017}} [arXiv:1801.06929].

\bibitem{BeltranJimenez:2017tkd}
J.~Beltr\'an Jim\'enez, L.~Heisenberg and T.~Koivisto, Coincident General Relativity, 
\href{https://journals.aps.org/prd/abstract/10.1103/PhysRevD.98.044048}{\textcolor{magenta}{Phys. Rev. D \textbf{98}, no.4, 044048 (2018)}} [arXiv:1710.03116]

\bibitem{BeltranJimenez:2018vdo}
J.~Beltr\'an Jim\'enez, L.~Heisenberg and T.~S.~Koivisto, Teleparallel Palatini theories, \href{https://iopscience.iop.org/article/10.1088/1475-7516/2018/08/039}{\textcolor{magenta}{ JCAP \textbf{08} (2018), 039 }} [arXiv:1803.10185]

\bibitem{Gomes:2022vrc}
D.~A.~Gomes, J.~Beltr\'an Jim\'enez and T.~S.~Koivisto, Energy and entropy in the geometrical trinity of gravity, \href{https://journals.aps.org/prd/abstract/10.1103/PhysRevD.107.024044}{\textcolor{magenta}{ Phys. Rev. D \textbf{107} (2023) no.2, 024044}} [arXiv:2205.09716]

\bibitem{Golovnev:2020zpv}
A.~Golovnev and M.~J.~Guzm\'an, Foundational issues in f(T) gravity theory, \href{https://www.worldscientific.com/doi/abs/10.1142/S0219887821400077}{\textcolor{magenta}{Int. J. Geom. Meth. Mod. Phys. \textbf{18} (2021) no.supp01, 2140007}} [arXiv:2012.14408].

\bibitem{Golovnev:2017dox}
A.~Golovnev, T.~Koivisto and M.~Sandstad, On the covariance of teleparallel gravity theories, \href{https://iopscience.iop.org/article/10.1088/1361-6382/aa7830}{\textcolor{magenta}{ Class. Quant. Grav. \textbf{34}, no.14, 145013 (2017)}} [arXiv:1701.06271]

\bibitem{Golovnev:2023yla}
A.~Golovnev, The geometrical meaning of the Weitzenb\"ock connection, \href{https://arxiv.org/abs/2302.13599}{\textcolor{magenta}{ [arXiv:2302.13599]}}.

\bibitem{Golovnev:2021omn}
A.~Golovnev and M.~J.~Guzman, Lorentz symmetries and primary constraints in covariant teleparallel gravity, 
\href{https://journals.aps.org/prd/abstract/10.1103/PhysRevD.104.124074}{\textcolor{magenta}{ Phys. Rev. D \textbf{104} (2021) no.12, 124074}} [arXiv:2110.11273].

\bibitem{Blixt:2022rpl}
D.~Blixt, R.~Ferraro, A.~Golovnev and M.~J.~Guzm\'an, Lorentz gauge-invariant variables in torsion-based theories of gravity, \href{https://journals.aps.org/prd/abstract/10.1103/PhysRevD.105.084029}{\textcolor{magenta}{ Phys. Rev. D \textbf{105} (2022) no.8, 084029}} [arXiv:2201.11102].

\bibitem{Golovnev:2022rui}
A.~Golovnev, On the Role of Constraints and Degrees of Freedom in the Hamiltonian Formalism, \href{https://www.mdpi.com/2218-1997/9/2/101}{\textcolor{magenta}{ Universe \textbf{9} (2023) no.2, 101}} [arXiv:2212.11260].

\bibitem{BoJe} C.~G.~Boehmer and E.~Jensko, Modified gravity: A unified approach, \href{https://journals.aps.org/prd/abstract/10.1103/PhysRevD.104.024010}{\textcolor{magenta}{ Phys. Rev. D \textbf{104} (2021) no.2, 024010}} [arXiv:2103.15906].

\bibitem{Boehmer:2023fyl}
C.~G.~Boehmer and E.~Jensko, Modified gravity: a unified approach to metric-affine models, \href{https://arxiv.org/abs/2301.11051}{\textcolor{magenta}{ [arXiv:2301.11051]}}.

\bibitem{BeltranJimenez:2022azb}
J.~Beltr\'an Jim\'enez and T.~S.~Koivisto, Lost in translation: The Abelian affine connection (in the coincident gauge), \href{https://www.worldscientific.com/doi/10.1142/S0219887822501080}{\textcolor{magenta}{ Int. J. Geom. Meth. Mod. Phys. \textbf{19} (2022) no.07, 2250108}} [arXiv:2202.01701]

\bibitem{Ferraro:2016wht}
R.~Ferraro and M.~J.~Guzm\'an, Hamiltonian formulation of teleparallel gravity, \href{https://journals.aps.org/prd/abstract/10.1103/PhysRevD.94.104045}{\textcolor{magenta}{Phys. Rev. D \textbf{94}, no.10, 104045 (2016)}} [arXiv:1609.06766]

\bibitem{Itin:2016nxk}
Y.~Itin, F.~W.~Hehl and Y.~N.~Obukhov, Premetric equivalent of general relativity: Teleparallelism, \href{https://journals.aps.org/prd/abstract/10.1103/PhysRevD.95.084020}{\textcolor{magenta}{Phys. Rev. D \textbf{95}, no.8, 084020 (2017)}} [arXiv:1611.05759]

\bibitem{Guzman:2020kgh}
M.~J.~Guzman and S.~Khaled Ibraheem, Classification of primary constraints for new general relativity in the premetric approach, \href{https://www.worldscientific.com/doi/abs/10.1142/S021988782140003X}{\textcolor{magenta}{ Int. J. Geom. Meth. Mod. Phys. \textbf{18}, no.supp01, 2140003 (2021)}} [arXiv:2009.13430]

\bibitem{BeltranJimenez:2019tme}
J.~Beltr\'an Jim\'enez, L.~Heisenberg, T.~S.~Koivisto and S.~Pekar, Cosmology in $f(Q)$ geometry, \href{https://doi.org/10.1103/PhysRevD.101.103507}{\textcolor{magenta}{ Phys. Rev. D \textbf{101}, no.10, 103507 (2020)} } [arXiv:1906.10027]

\bibitem{LorGov} A. Golovnev, Issues of Lorentz-invariance in f(T) gravity and calculations for spherically symmetric solutions, \href{https://iopscience.iop.org/article/10.1088/1361-6382/ac2136}{\textcolor{magenta}{ Classical and Quantum Gravity {\bf 38} (2021) 197001}} [arXiv:2105.08586]

\bibitem{unieq} G.R. Bengochea, G. Leon, A. Perez, and D. Sudarsky, A clarification on prevailing misconceptions in unimodular gravity, arXiv:2308.07360

\bibitem{unipert} C. Gao, R.H. Brandenberger, Y. Cai, and P. Chen, Cosmological Perturbations in Unimodular Gravity, JCAP09(2014)021 [arXiv:1405.1644]

\bibitem{BeltranJimenez:2019esp}
J.~Beltr\'an Jim\'enez, L.~Heisenberg and T.~S.~Koivisto, The Geometrical Trinity of Gravity, \href{https://doi.org/10.3390/universe5070173}{\textcolor{magenta}{Universe \textbf{5}, no.7, 173 (2019)}}

\bibitem{GSVnewGR} A. Golovnev, A.N. Semenova, and V.P. Vandeev, Static spherically symmetric solutions in New General Relativity, \href{https://arxiv.org/abs/2305.03420}{\textcolor{magenta}{[arXiv:2305.03420]}}

\bibitem{Lavinia} F. D'Ambrosio, L. Heisenberg, and S. Zentarra, Hamiltonian Analysis of f(Q) Gravity and the Failure of the Dirac-Bergmann Algorithm for Teleparallel Theories of Gravity, arXiv:2308.02250

\bibitem{Rastall} P. Rastall, Generalization of the Einstein theory, \href{https://journals.aps.org/prd/abstract/10.1103/PhysRevD.6.3357}{\textcolor{magenta}{Phys. Rev. D {\bf 6} (1972) 3357}} 

\bibitem{Visser} M. Visser, Rastall gravity is equivalent to Einstein gravity, \href{https://www.sciencedirect.com/science/article/pii/S0370269318303927}{\textcolor{magenta}{ Phys. Lett. B {\bf 782} (2018) 83}} [arXiv:1711.11500]

\bibitem{GaR} A. Golovnev, More on the fact that Rastall = GR, Annals of Physics 461 (2024) 169580 [arXiv:2311.00131]

\bibitem{Salva} S. Capozziello, M. Capriolo, and Sh. Nojiri, Gravitational waves in f(Q) non-metric gravity via geodesic deviation, arXiv:2401.06424

\bibitem{CairoBH} A. Awad, A. Golovnev, M.J. Guzman and W. El Hanafy, Revisiting diagonal tetrads: New Black Hole solutions in f(T) gravity, \href{https://link.springer.com/article/10.1140/epjc/s10052-022-10939-0}{\textcolor{magenta}{ Eur. Phys. J. C {\bf 82} (2022) 972}} [arXiv:2207.00059]

\bibitem{De:2022jvo}
A.~De and T.~H.~Loo, On the viability of f(Q) gravity models, \href{https://iopscience.iop.org/article/10.1088/1361-6382/accef7}{\textcolor{magenta}{ Class. Quant. Grav. \textbf{40} (2023) no.11, 115007}}
[arXiv:2212.08304].

\bibitem{Dimakis:2022rkd}
N.~Dimakis, A.~Paliathanasis, M.~Roumeliotis and T.~Christodoulakis, FLRW solutions in f(Q) theory: The effect of using different connections, \href{https://journals.aps.org/prd/abstract/10.1103/PhysRevD.106.043509}{\textcolor{magenta}{ Phys. Rev. D \textbf{106} (2022) no.4, 043509}} [arXiv:2205.04680].

\bibitem{Paliathanasis:2023nkb}
A.~Paliathanasis, Dynamical analysis of $f(Q)$-cosmology, \href{https://www.sciencedirect.com/science/article/pii/S2212686423000894}{\textcolor{magenta}{   Phys. Dark Univ. \textbf{41} (2023), 101255}} [arXiv:2304.04219].

\bibitem{Golovnev:2020nln}
A.~Golovnev and M.~J.~Guzman, Nontrivial Minkowski backgrounds in $f(T)$ gravity, \href{https://journals.aps.org/prd/abstract/10.1103/PhysRevD.103.044009}{\textcolor{magenta}{
Phys. Rev. D \textbf{103} (2021) no.4, 044009
doi:10.1103/PhysRevD.103.044009}}
[arXiv:2012.00696].

\bibitem{Bahamonde:2022zgj}
S.~Bahamonde and L.~J\"arv, Coincident gauge for static spherical field configurations in symmetric teleparallel gravity, \href{https://link.springer.com/article/10.1140/epjc/s10052-022-10922-9}{\textcolor{magenta}{ Eur. Phys. J. C \textbf{82}, no.10, 963 (2022)} } [arXiv:2208.01872]

\bibitem{Ferraro:2020tqk}
R.~Ferraro and M.~J.~Guzm\'an, Pseudoinvariance and the extra degree of freedom in f(T) gravity, \href{https://journals.aps.org/prd/abstract/10.1103/PhysRevD.101.084017}{\textcolor{magenta}{   
Phys. Rev. D \textbf{101} (2020) no.8, 084017}}
[arXiv:2001.08137].

\end{thebibliography}
\end{document}